\documentclass[12pt]{article}

\usepackage{amsmath}
\usepackage{graphicx}
\usepackage{enumerate}
\usepackage{natbib}

\usepackage{url} 
\usepackage{amssymb,amsfonts}
\usepackage[all,arc]{xy}
\usepackage{mathrsfs}
\usepackage{mathtools}
\usepackage{lipsum}
\usepackage{changepage}
\usepackage{upgreek}
\usepackage{bm}
\usepackage{tabu}
\usepackage{chngcntr}
\usepackage{hyperref}
\usepackage{cleveref}
\hypersetup{colorlinks, citecolor=blue}
\usepackage{listings}

\newcommand{\blind}{0}

\begin{document}

\newtheorem{theorem}{Theorem}
\newtheorem{lemma}{Lemma}
\newtheorem{corollary}{Corollary}

\def\spacingset#1{\renewcommand{\baselinestretch}%
{#1}\small\normalsize} \spacingset{1}


\if1\blind
{
  \title{\bf Revisiting the probabilistic method of record linkage}
  \author{Author 1\thanks{
    The authors gratefully acknowledge \textit{please remember to list all relevant funding sources in the unblinded version}}\hspace{.2cm}\\
    Department of YYY, University of XXX\\
    and \\
    Author 2 \\
    Department of ZZZ, University of WWW}
  \maketitle
} \fi


\title{\bf Revisiting the probabilistic method of record linkage}
\author{A. Dasylva\footnote{abel.dasylva@canada.ca}, A. Goussanou, D. Ajavon\\
Statistics Canada\\
and \\
H. Abousaleh\\
Department of Mathematics and Statistics, University of Victoria}
\maketitle

\begin{abstract}
\noindent In theory, the probabilistic linkage method provides two distinct advantages over nonprobabilistic methods, including minimal rates of linkage error and accurate measures of these rates for data users.
 However, implementations can fall short of these expectations
 either because the {conditional independence} assumption is made, or because a model with interactions is used but lacks the {identification} property.
 In official statistics, this is currently the main challenge to
 the automated production and use of linked data.
 To address this challenge, a new methodology is described for {\it proper} linkage problems, where matched records may be identified with a probability that is {bounded away from zero}, regardless of the population size.
It models the number of {\it neighbours} of a given record, i.e. the number of resembling records.
 To be specific, the proposed model is a finite mixture where each component is the  sum of a Bernoulli variable with an independent Poisson variable.
It has the identification property and yields solutions for many longstanding problems, including the evaluation of {blocking criteria} and the estimation of linkage errors for probabilistic or {nonprobabilistic} linkages, all without clerical reviews or
 conditional independence assumptions.
Thus it also enables unsupervised machine learning solutions for record linkage problems.
\end{abstract}


\noindent%


\vspace{12pt}

\noindent {\bf Disclaimer}:
The content of this article represents the position of the authors and may not necessarily represent that of Statistics Canada. This article describes theoretical approaches and does not reflect currently implemented methods at Statistics Canada.

\vfill

\newpage
\spacingset{1.5} 
\section{Introduction}
\label{sec:intro}

\noindent Fifty years ago, Fellegi and Sunter published their landmark paper \cite{fellegi_sunter_1969}, where they formalized probabilistic record linkage.
Since then, record linkage has become an important tool in official statistics.
It consists in finding records from the {\it same} individual in different files and differs from {\it statistical matching}
 \cite{dorazio_et_al_2006}; a form of imputation which consists in finding records from {\it similar} individuals.
Record linkage is also different from {\it automated coding}, where the goal is assigning a given observation to a class among  a fixed number of classes.

\noindent There are countless competing solutions for implementing a record linkage \cite{christen_2012}.
Yet in the probabilistic method, errors play a central role.
They include false positives and false negatives, where a false positive is a link between two unmatched records and a false negative is the absence of a link between two matched records.
In its fully automated form (i.e. with a single weight threshold) the probabilistic method is simply an application of Neyman-Pearson lemma \cite{neyman_pearson_1933}.
 Hence it promises two major advantages over its competitors.
%
%
The first advantage is a minimal type II error for a fixed type I error, under the null hypothesis that two records are matched.
Alternatively, the linkage can minimize the type I error for a fixed type II error.
The second advantage is the accurate estimation of both kinds of error in a fully automated manner.
In practice, the promise of the probabilistic method has been elusive, because fulfilling it requires a very good statistical model for the estimation of the necessary linkage parameters, including the linkage weights and the related thresholds.
The most common model has been a log-linear mixture with the assumption that the linkage variables are conditionally independent given a pair match status \cite{fellegi_sunter_1969}.
However, this assumption has lead to biased estimators because, quite often, it has been inconsistent with actual data.
For example, when linking social data, given names and last names  are not independent for a given individual.
To overcome this limitation, log-linear mixtures with interactions have been considered.
Yet, these models have been hard to use.
Indeed, in the past fifty years, they have been successfully applied without labeled pairs (i.e. pairs that are known to be matched or unmatched), but only in a handful of studies including  that by Winkler \cite{winkler_1993}.
Since log-linear models with interactions are not generally known to have the identification property, some of them may lack it.
In such a model, there are simply too many parameters to estimate from the data at hand, no matter how large the sample.
This a serious problem for analysts, who need accurate measures of the error rates \cite{neter_maynes_ramanathan_1965, scheuren_winkler_1993,  ding_fienberg_1994, lahiri_larsen_2005, chipperfield_bishop_campbell_2011, kim_chambers_2016, hof_and_zwinderman_2015}.
Besides, proving the identification of a log-linear mixture model is a non-trivial matter, even when the conditional assumption is made \cite{fienberg_2009}.
%
%
 In this regard, having fewer model parameters than estimating equations falls short of
 proving the identification property \cite{fienberg_2009}.


\noindent Probabilistic linkage suffers from another limitation regarding false negatives caused by blocking criteria.
Indeed, blocking is an essential operation when linking large files.
Yet it generates false negatives that are hard to measure given the sheer size of the Cartesian product.
Herzog et al. \cite{herzog_scheuren_winkler_2007} have described a capture-recapture technique that requires two conditionally independent blocking keys.
 Their method counts the number of matched pairs that agree
 on the first key, on the second key and on both.
 In practice, applying this method is challenging because
 there are  usually no such keys.
Even if it were the case, the method requires the identification of each matched pair that meets a given blocking criterion.
With large files that include millions of records, this is   prohibitively expensive even if resources for reliable clerical reviews are available.

\noindent The remaining sections are organized as follows.
%
%
Section~\ref{section: notation and assumptions} presents the notations and assumptions.
It also introduces the crucial concepts of {\it neighbourhood} and {\it proper} linkage problem.
Section~\ref{section: linkage parameters and errors} articulates the relationship between the probabilistic linkage parameters and the error rates, on one hand, and the distributional parameters for the number of neighbours, on the other hand.
Section~\ref{section: estimation} describes the estimation procedure for the distribution of the number of neighbours.
It involves an Expectation-Maximization (E-M) procedure \cite{{dlr_1977}}.
Section~\ref{section: simulations} covers the simulations.
Section~\ref{section: empirical study} describes the empirical study.
Section~\ref{section: future work} concludes with future work.
All the proofs are found in the supplementary material.


\section{Notation and assumptions}\label{section: notation and assumptions}

\noindent In probabilistic record linkage, statistical inference about linkage parameters and errors is commonly based on the maximization of a likelihood for a sample of record pairs \cite{winkler_1988,jaro_1989}.
 Most of the time, this likelihood is actually a
 {\it composite} likelihood \cite{varin_et_al_sinica_2011}
 because the pairs are correlated.
 In general, the choice of maximum likelihood procedures is
 justified by the asymptotic properties of the resulting
 estimators when the sample size tends to infinity.
 In our context, this makes sense if record linkage remains
 a viable option even when the population becomes arbitrarily large.
 This is possible only if the linkage variables provide more information as the population grows, such that matched records may be identified with a probability that is bounded away from zero, regardless of the population size.
 To discuss procedures for valid statistical inference about
 linkage parameters or errors, we restrict ourselves to the class
 of {\it proper} linkage problems, which satisfy this
 property and include all practical situations where record
 linkage is a viable alternative.
 Outside this class, problems are {\it improper} and better
 viewed as satistical matching or automated coding problems.
 The concept of a proper linkage problem is also consistent with
 the use of the Shanon entropy
 \cite[chap. 2.1]{cover_thomas_1991}
 to measure the discriminating power of linkage variables and
 to decide if two files could
 be linked \cite[chap. 4.2]{rlppm_2017}.
 Discussing what might happen to a linkage when the
 population size grows is also crucial for
 the future of official statistics.
 Indeed, many national statistical offices have recently
 embarked on the construction of an
 infrastucture of linked statistical registers
 \cite{wallgren_2014}, through multiple linkages,
 including many without a unique identifier.
 This discussion can provide important insights into current
 design decisions that will have a major impact on the production
 of official statistics and public policy for decades to come.
 We next give some notation and a mathematical definition of
 what constitutes a proper linkage problem,
 including important consequences of this definition.


\noindent {\it Notation}:
With few exceptions, we denote a scalar by a lower-case letter and a vector or matrix by  a boldface lower-case letter.
This convention also applies to a random variable, vector or matrix.
In general, the same symbol is used for a random variable (vector or matrix) and for a specific value of this variable (vector or matrix).
However, when needed for clarity, an upper-case letter is used for a random variable (vector or matrix) while the corresponding lower-case letter is used for a specific value.
In general, a scalar function is denoted by a lower-case letter while a vector-valued (or matrix-valued) function is denoted by a boldface lower-case letter.
%
%
For a twice differentiable scalar multivariate function $f(.)$ of $\bm{\psi} = \left ( \psi_1,\ldots,\psi_q\right ) \in I\!\!R^q$, we let
 $\left ( \partial/ \partial \bm{\psi}^{\top}\right ) f =
  \partial f/ \partial \bm{\psi}^{\top} =
  \left [ \partial f / \partial \psi_1 \ldots
  \partial f / \partial \psi_q  \right ]$ \cite{schott_1997}.
 We also denote the Hessian matrix of $f(.)$ by
 $\left ( \partial^2/ \partial \bm{\psi}
  \partial \bm{\psi}^{\top}\right ) f =
  \left [ \partial^2 f / \partial \psi_r \partial \psi_s  \right ]_{1 \leq r,s \leq q}$.

\noindent For our specific problem, consider a large population of $N$ independent and identically distributed individuals.
Individual $i$ is associated with record $\bm{v}_i$ in register $A$ and record $\bm{v}_{\pi(i)}'$ in register $B$, where $\bm{v}_i$ and $\bm{v}_{\pi(i)}'$ belong to a set ${\cal V}_N$ of (possibly high-dimensional) categorical vectors and  $\pi(.)$ is an unknown random permutation of $\{1,\ldots,N\}$.
The permutation is assumed to be uniformly random independently of the records, i.e.
%
%
\begin{equation}
P\left ( \pi(.) \left |  \left [ \left ( \bm{v}_i, \bm{v}_{\pi(i)}' \right ) \right ]_{1\leq i \leq N} \right  . \right ) = \frac{1}{N!}
\end{equation}
The sample $\left [ \left ( \bm{v}_i, \bm{v}_{\pi(i)}' \right ) \right ]_{1 \leq i \leq N}$ is independent and identically distributed.
Let $M=\{ j=\pi(i)\}$ denote the event that the pair $(i,j)$ is matched, and $U=\{ j\neq \pi(i)\}$ the event that it is unmatched.
Note that what is to follow also applies if register A is instead a simple random sample.


\noindent {\it Neighbourhoods}:
 It is a collection
 $\left [ {\cal B}_N \left ( \bm{v} \right ) \right ]_{\bm{v} \in {\cal V}_N}$ of subsets of
 ${\cal V}_N$ such that $\bm{v} \in {\cal B}_N (\bm{v})$ for each $\bm{v}$. 
 Record $\bm{v}_j'$ is a {\it neighbour} of record $\bm{v}_i$ if
 $\bm{v}_j' \in {\cal B}_N \left ( \bm{v}_i \right )$. 
 Define the {\it number of neighbours} of record $i$ by {\small $n_i = \sum_{j=1}^N I \left ( \bm{v}_j' \in {\cal B}_N \left ( \bm{v}_i \right ) \right )$}, where
 {\small $n_i=n_{i|M}+n_{i|U}$};
{\small $n_{i|M}=I \left ( \bm{v}_{\pi(i)}' \in {\cal B}_N \left ( \bm{v}_i \right ) \right )$} being the number of {\it matched} neighbours and {\small $n_{i|U}=\sum_{i'\neq i} I \left ( \bm{v}_{\pi(i')}' \in {\cal B}_N \left ( \bm{v}_i \right ) \right )$} being the number of {\it unmatched} neighbours.
 The collection
 {\small $\left [ {\cal B}_N(\bm{v})\right ]_{\bm{v} \in {\cal V}_N}$} may
 be based on blocking criteria, comparison outcomes (including  
 blocking criteria) or a linkage decision based on a
 nonprobabilistic method, provided this latter decision only
 depends on the two records in the related pair.
 This latter condition means that the nonprobabilistic linkage may be described
 by a mathematical map\footnote{This is different from the concept of {\it mapping};
 another name for conflict resolution in record linkage.
 A conflict occurs if a record is matched with at most one record but linked to many records,
 e.g., when linking individuals to death events.
 It is resolved by keeping a single candidate link.
 With mapping, the linkage {\it cannot} be described as a mathematical map from
 ${\cal V}_N \times {\cal V}_N$ into $\{0,1\}$.}
 $L: {\cal V}_N \times {\cal V}_N \mapsto \{0,1\}$.
 For given blocking criteria, define
 ${\cal B}_N \left ( \bm{v} \right )$ as the set of all
 $\bm{v}' \in {\cal V}_N$ such that the pair $(\bm{v},\bm{v}')$
 satisfies the criteria.
 We can also define the collection based on a
 given vector of comparison outcomes.
 This vector encodes the information about the comparison of the
 different variables.
 It is a vector of dichotomous outcomes if comparisons are based
 on full agreement.
 Let $\gamma (\bm{v},\bm{v}')$ denote the comparison vector for
 the pair $(\bm{v},\bm{v}')$.
 In this case, the collection  ${\cal B}_N(\bm{v})$ may include
 each $\bm{v}'$ such that $\gamma (\bm{v},\bm{v}')$ is 
 equal to a specific value, e.g., full agreement on each
 variable.
 Finally, for a nonprobabilistic linkage (that is equivalent to a map as mentioned above)
 , we can let ${\cal B}_N (\bm{v})= \left \{ \bm{v}' \in {\cal V}_N \
 s.t. \ L(\bm{v},\bm{v}')=1 \right \}$.

%
\noindent {\it Unmatched neighbours and frequency weight}:
 In probabilistic linkage, frequency weights are used to leverage the fact that two
 records are more likely to be matched if they agree on a rare value than on a more
 common one \cite{fellegi_sunter_1969}.
 The number of unmatched neighbours $n_{i|U}$ is related to this frequency weight.
 Indeed it provides a measure of the density of records at $\bm{v}_i$.
 Thus  $-\log \left ( E \left [ \left . n_{i|U}/N \right | \bm{v}_i\right ] \right )$
 may be viewed as a smoothed version of the frequency weight,
 which would be used if comparing the records based on a full agreement on all the variables.
%
%
 This smoothed frequency weight is based on the average density of records in the neighbourhood
 ${\cal B}_N \left ( \bm{v}_i \right )$.
 It addresses a current practical problem when applying the probabilistic method.
 Indeed, frequency weights are most beneficial when applied to infrequent values.
 Estimating the frequencies of such values is inherently difficult
 and requires some smoothing.

%
\noindent {\it Proper linkage problem}:
 For a collection $\left [ {\cal B}_N(\bm{v})\right ]_{\bm{v} \in {\cal V}_N}$ of
 subsets of ${\cal V}_N$, define the probabilities
%
%
\begin{eqnarray}
p_N \left ( \bm{v}_i \right ) &=& P \left ( \left . \bm{v}_{\pi(i)}' \in {\cal B}_N \left ( \bm{v}_i \right ) \right | \bm{v}_i \right ), \\[6pt]
\lambda_N \left ( \bm{v}_i \right ) &=& P \left ( \left . \bm{v}_{\pi(i')}' \in {\cal B}_N \left ( \bm{v}_i \right ) \right | \bm{v}_i \right ), \ i' \neq i.
\end{eqnarray}
 The problem is {\it proper} if there exists a
 collection $\left [ {\cal B}_N(\bm{v})\right ]_{\bm{v} \in {\cal V}_N}$ of
 subsets of ${\cal V}_N$, such that the following conditions apply:
%
%
\begin{eqnarray}
\label{eq: very first condition}
\inf_{\bm{v} \in {\cal V}_N} p_N \left ( \bm{v} \right )  &\geq& \delta, \\[6pt]
\label{eq: first condition}
\sup_{\bm{v} \in {\cal V}_N} (N-1) \lambda_N \left ( \bm{v} \right ) 
 &\leq& \Lambda, \\[6pt]
\label{eq: second condition}
\left ( p_N(\bm{v}_i), (N-1) \lambda_N \left ( \bm{v}_i \right ) \right )
&\sim& F(.,.),
\end{eqnarray}
 where $\delta>0$ and $F(.,.)$ does not depend on $N$.

\noindent In a proper linkage problem, the expected number of neighbours is bounded above (by $1+\Lambda$) regardless of the population size.
 When the set ${\cal B}_N (\bm{v})$ is based on blocking criteria, the upperbound also implies that there are $O(N)$  pairs that satisfy the blocking criteria.


\noindent Eqs.~(\ref{eq: first condition}) and (\ref{eq: second condition}) imply that the matched record of a given record may be identified with a probability that is bounded away from zero.
Indeed, for record $\bm{v}_i$, consider the procedure that attemps to identify record $\bm{v}_{\pi(i)}'$ as follows.
It fails if ${\cal B}_N (\bm{v}_i)$ is empty.
 Otherwise it randomly selects a neighbour.
Let $\tau=$ denote the success probability of this procedure.
 It is bounded below as follows (see the proof in the supplementary material.)
%
%
\begin{equation}
\label{eq success probability lower bound}
\tau \geq \frac{\delta}{1+\Lambda}.
\end{equation}

%
\noindent {\it Shannon entropy}: 
 With social data, the family name is an important linkage variable that is
 characterized by a power law frequency distribution \cite{fox_lasker_1983}, where
 a few values are much more frequent than most other values.
 Such a variable may be combined with other variables, e.g. the birthdate and the address,
%
 to produce records that meet Eqs.~(\ref{eq: first condition}) and (\ref{eq: second condition}).
%
%
 This condition also implies a lower bound on the Shannon entropy  $H_N$ (\cite[chap. 2.1]{cover_thomas_1991}) of the record $\bm{v}_j'$, if there are $\bm{v}_{(1)},\ldots,\bm{v}_{(R_N)}$ from ${\cal V}_N$ such that
 ${\cal B}_N \left ( \bm{v}_{(1)} \right ),\ldots,
 {\cal B}_N \left ( \bm{v}_{(R_N)} \right )$ are disjoint and
 {\small $P \left ( \bm{v}_j' \in
 \bigcup_{r=1}^{R_N} {\cal B}_N \left ( \bm{v}_{(r)}\right ) \right )=
 1-O \left ( N^{-1}\right )$}.
 Indeed, we have (see the proof in the supplementary material)
%
%
\begin{equation}
\label{eq shannon entropy lower bound}
H_N \geq \log(N-1) -\log \Lambda +
 O \left ( N^{-1}\log N\right ).
\end{equation}


\section{Linkage parameters and errors}
\label{section: linkage parameters and errors}

\noindent This section expands on the relationship between the joint distribution $F(.,.)$ on one hand and the probabilistic linkage parameters and the rates of linkage errors on the other hand.
 All proofs and derivations are found in the supplementary material.

%
\noindent The m-probability is the conditional probability that two matched records are neighbours.
It is given by
%
%
\begin{equation}
\label{eq m-probability}
P \left ( \left . \bm{v}_j' \in {\cal B}_N \left ( \bm{v}_i \right ) \right | M \right ) =
E[p].
\end{equation}
%
%
The u-probability is the conditional probability that two unmatched records are neighbours.
It is given by
%
%
\begin{equation}
\label{eq u-probability}
 P \left ( \left . \bm{v}_j' \in {\cal B}_N \left ( \bm{v}_i 
 \right ) \right | U \right )
=
\frac{E[\lambda]}{N-1}.
\end{equation}
%
%
Consequently, the linkage weight of the event $\left \{ \bm{v}_j' \in {\cal B}_{N} \left ( \bm{v}_i\right ) \right \}$ is
%
%
\begin{equation}
\label{eq linkage weight}
w =
\log \left ( \frac{(N-1)E[p]}{E[\lambda]}\right ).
\end{equation}
%
%
Finally, the conditional match probability of the event $\bm{v}_j' \in {\cal B}_{N} \left ( \bm{v}_i\right )$ is
%
%
\begin{equation}
\label{eq conditional match probability}
P \left ( M \left | \bm{v}_j' \in {\cal B}_{N} \left ( \bm{v}_i\right ) \right . \right ) =
 \left ( 1 + \frac{E[\lambda]}{E[p]} \right )^{-1}.
\end{equation}

%
\noindent Measures of linkage accuracy may be provided at the pair level or at the record level but they are usually provided at the former level.
If classical models are limited to the pair level, the proposed methodology does not have this limitation as demonstrated by Eq.~(\ref{eq: linkage accuracy at record level}).
With this latter methodology, the pairwise rates are computed as follows.
A distinct collection $\left [ {\cal B}_N(\bm{v})\right]_{\bm{v} \in {\cal V}_N}$ is used for each value of the comparison vector, with the corresponding m-probability, u-probability, linkage weight and conditional match probability. 
%
%
These parameters are then applied to estimate the error rates, including the False Negative Rate (FNR) and the False Positive Rate (FPR), as prescribed by Fellegi and Sunter \cite{fellegi_sunter_1969}, where the FPR is instead called False Match Rate (FMR).
The same formulas may be used to estimate the linkage errors for blocking criteria and  nonprobabilistic linkages (that correspond to a map), by defining the collection of 
neighbourhoods as described in the previous section.

%
\noindent The proposed model also yields estimates of linkage error at the record level, e.g. the conditional Probability that the matched record is among the neighbours of record $i$.
When $n_i$ is positive, this probability is given by
%
%
\begin{equation}
\label{eq: linkage accuracy at record level}
P \left ( n_{i|M}=1 \left | n_i \right . \right ) \approx
\frac{E \left [ p e^{-\lambda} \frac{\lambda^{n_i-1}}{(n_i-1)!}  \right ]}{ E \left [ e^{-\lambda} \frac{\lambda^{n_i-1}}{(n_i-1)!} \left ( p + (1-p) \frac{\lambda}{n_i}  \right ) \right ]},
\end{equation}
where $E \left [ p e^{-\lambda} \frac{\lambda^{n_i-1}}{(n_i-1)!}  \right ] = \int p e^{-\lambda} \frac{\lambda^{n_i-1}}{(n_i-1)!}  dF$, and
$E \left [ (1-p) e^{-\lambda} \frac{\lambda^{n_i}}{n_i!} \right ]=\int (1-p) e^{-\lambda} \frac{\lambda^{n_i}}{n_i!} dF$.
In the above formula, the right-hand side no longer depends on the population size.
It shows that we may have a false positive even if there is a single neighbour.
Note that record-level estimates of linkage errors are difficult to obtain with previous models that focus on a single record pair.

%
\noindent The above discussion shows that, in a proper linkage problem,
 the probabilistic linkage parameters and the rates of error are
 fully characterized by the joint distribution $F(.,.)$.
 The next section looks at the estimation of this key parameter.


\section{Estimation}\label{section: estimation}

\noindent In probabilistic record linkage, the estimation of linkage parameters has been commonly based on log-linear mixtures and the maximization of a composite likelihood, for the sample of record pairs that meet the blocking criteria.
In this procedure, one usually views the variable sample size as a fixed quantity.
%
%
This also means that one ignores any statistical information that comes with this variable or with other related variables, such as the number of neighbours of different records.
Instead, the estimation relies entirely on the observed proportions of pairs, for different combinations of comparison outcomes.

\noindent In what follows, the distribution of neighbours is shown to provide the 
information for estimating the parameter $F(.,.)$ and thus the linkage parameters as well as  the errors.
 To be specific, when the distribution $F(.,.)$ is discrete and the population size
 is large, the number of neighbours follows a finite mixture model that has the
 identification property.
 In this mixture, each component is the sum of a Bernoulli variable with
 an independent Poisson variable.
 The proposed estimation procedure involves an Expectation-Maximization (E-M) algorithm
 that is applied to a carefully selected sample of {\it asymptotically independent}
 observations.
 All proofs and derivations are found in the supplementary material.


\subsection{A finite mixture model}

\noindent In what follows, consider that $F(.,.)$ is (well approximated by)  a discrete CDF  of the form $\sum_{g=1}^G \alpha_g I \left ( p_g \leq p, \lambda_g \leq \lambda \right )$.
To motivate the model that is to be proposed for $n_i$, first note that for any
 $(p,\lambda) \in (\delta,1)\times(0, \Lambda)$, we have
 $$n_{i|U} \left |
 \left \{ \left ( p_N(\bm{v}_i),(N-1) \lambda_N(\bm{v}_i) \right )=(p,\lambda)\right \} \right . \sim
 Binomial(N-1,\lambda/(N-1)).$$
 Since $Binomial(N-1,\lambda/(N-1)) \stackrel{d}{\rightarrow} Poisson(\lambda)$ \cite[Theorem 23.2]{billingsley_1995},
 when $N \rightarrow \infty$, $n_i=n_{i|M}+n_{i|U}$ where
 $n_{i|M}$ and $n_{i|U}$ are conditionally independent given
 $\left ( p_N(\bm{v}_i),(N-1) \lambda_N(\bm{v}_i) \right )$ and
%
%
\begin{eqnarray}
\left . {n}_{i|M} \right | \left \{ \left ( p_N(\bm{v}_i),(N-1) \lambda_N(\bm{v}_i) \right )=(p_g,\lambda_g) \right \} &\sim&
{Bernoulli}(p_g), \\[6pt]
\left . {n}_{i|U} \right | \left \{ \left ( p_N(\bm{v}_i),(N-1) \lambda_N(\bm{v}_i) \right )=(p_g,\lambda_g) \right \}
 &\dot{\sim}& {Poisson}(\lambda_g).
\end{eqnarray}
 Then
%
%
\begin{equation}
 P \left ( n_i=k \right ) \dot{=}
 \sum_{g=1}^G \alpha_g \left ( \left (1-p_g\right )e^{-\lambda_g} I(k=0)+
 I(k>0) \frac{e^{-\lambda_g} \lambda_g^{k-1}}{(k-1)!}
 \left ( p_g + \left ( 1-p_g\right ) \frac{\lambda_g}{k}\right ) \right )
\end{equation}
 The following theorem examines the identification of this model.
%

\begin{theorem}\label{theorem: identifiability}
Consider $Y=Y_1+Y_2$, where $\left (Y_1,Y_2 \right )$ is distributed according to the mixture
%
%
\begin{equation}
 P \left ( Y_1=y_1, Y_2=y_2 \right ) =
 \sum_{g=1}^G \alpha_g p_g^{y_1} \left ( 1 - p_g \right )^{1-y_1}
 \frac{e^{-\lambda_g} \lambda_g^{y_2}}{y_2!}
\end{equation}
 Let $\lambda_{(1)}> \lambda_{(2)}>\ldots >\lambda_{\left ( G_0 \right )}$ denote
 the distinct $\lambda_g$ values, and define
%
%
\begin{eqnarray}
 \alpha_{(g)} &=&
 \sum_{g'=1}^G I \left ( \lambda_{g'} = \lambda_{(g)}\right ) \alpha_{g'} \\
 p_{(g)} &=&
 \frac{\sum_{g'=1}^G
 I \left ( \lambda_{g'} = \lambda_{(g)}\right )
 \alpha_{g'} p_{g'} }{\sum_{g'=1}^G I \left ( \lambda_{g'} = \lambda_{(g)}\right ) \alpha_{g'}}
\end{eqnarray}
 for $g=1,\ldots,G_0 \leq G$.
%
%
 Consider $Y'$ from the same parametric family, where $G_0'\leq G'$,
 {\footnotesize
 $\left [ \left ( \alpha_g', p_g', \lambda_g' \right ) \right ]_{1 \leq g \leq G'}$}, and {\footnotesize
 $\left [ \left ( \alpha_{(g)}', p_{(g)}', \lambda_{(g)}' \right )
 \right ]_{1 \leq g \leq G_0'}$} are the corresponding parameters.
 Then $Y$ and $Y'$ have the same distribution if and only if $G_0'=G_0$, and
%
%
\begin{eqnarray}
\lambda_{(g)}' &=& \lambda_{(g)}, \\[6pt]
\alpha_{(g)}' &=& \alpha_{(g)}, \\[6pt]
p_{(g)}' &=& p_{(g)},
\end{eqnarray}
 for $g=1,\ldots,G_0$.
\end{theorem}
 The above theorem suggests that we rewrite the probability $P \left ( n_i=k \right )$ as
%
%
\begin{equation}
 P \left ( n_i=k \right ) \approx
 \sum_{g=1}^{G_0} \alpha_{(g)} \left ( \left (1-p_{(g)}\right )e^{-\lambda_{(g)}} I(k=0)+
 I(k>0) \frac{e^{-\lambda_{(g)}} \lambda_{(g)}^{k-1}}{(k-1)!}
 \left ( p_{(g)} + \left ( 1-p_{(g)}\right ) \frac{\lambda_{(g)}}{k}\right ) \right )
\end{equation}
 According to the theorem, $G_0$ and
 $\left [ \left ( \alpha_{(g)}, p_{(g)}, \lambda_{(g)}\right ) \right ]$ are uniquely
 determined by the limiting distribution of $n_i$.
 In turn, these parameters determine the linkage parameters and errors through
 $E \left [ p \right ] = \sum_{g=1}^G \alpha_g p_g =
 \sum_{g=1}^G \alpha_{(g)} p_{(g)}$ and
 $E \left [ \lambda \right ] = \sum_{g=1}^G \alpha_g \lambda_g =
 \sum_{g=1}^G \alpha_{(g)} \lambda_{(g)}$.
%
%
 From now on, without losing any generality, we suppose that $G_0=G$ and let
 $\bm{\psi} = \left [ \left ( \alpha_{(g)}, p_{(g)},
 \lambda_{(g)} \right )\right ]_{1 \leq g \leq G}$ denote the vector of parameters.


\subsection{Asymptotic independence} \label{subsection: Asymptotic independence}

\noindent For statistical inference about the parameter vector $\bm{\psi}$, it is convenient to treat the observations $n_1,\ldots,n_M$ as if they were
 independent and identically distributed.
However, the estimated variances, confidence intervals and critical levels would be inaccurate.

\noindent In what follows, we show that this problem does not arise if the inference is instead based on a size $m_N$ simple random sample (without replacement) of observations,
 where $m_N\rightarrow \infty$ and $m_N = o \left ( \sqrt{N}\right )$.
 Without losing any generality (because the permutation $\pi(.)$ is uniformly random),
 we let this sample be $n_1,\ldots,n_{m_N}$.
 We also make the following further assumptions.
%
%
\begin{equation}
 \label{eq bound on disjointness probability}
 (N-1) P \left ( {\cal B}_N \left ( \bm{v}_i \right ) \cap
 {\cal B}_N \left ( \bm{v}_{i'} \right )  \neq \emptyset \right ) \leq \Lambda,
 \ i \neq i'
\end{equation}
 For $\bm{v}_i$ and $\bm{v}_{i'}$ such that
 ${\cal B}_N \left ( \bm{v}_i \right ) \cap
 {\cal B}_N \left ( \bm{v}_{i'} \right ) = \emptyset$, we have
%
%
\begin{equation}
 \label{eq bound on other capture probability}
 (N-1) P \left ( \left . \bm{v}_{\pi(i')}' \in {\cal B}_N \left ( \bm{v}_i \right ) \right |
 \bm{v}_i, \bm{v}_{i'} \right ) \leq \Lambda
\end{equation}

\noindent In the following paragraphs, we give
 Theorem~\ref{theorem asymptotic independence} and
 Corollary~\ref{corollary theorem asymptotic independence} and discuss their consequences
 regarding the consistency and asymptotic normality of the proposed estimator.
 For notational convenience let
 $\bm{p}_N = \left [ p_N \left ( \bm{v}_i \right )\right ]_{1 \leq i \leq m_N}$,
 $\bm{\lambda}_N = \left [ \lambda_N \left ( \bm{v}_i \right )\right ]_{1 \leq i \leq m_N}$
 and $\bm{n} = \left [ n_i \right ]_{1 \leq i \leq m_N}$.
%
%
\begin{theorem} \label{theorem asymptotic independence}

\noindent Let
 $\tilde{\tilde{\bm{n}}}_{M}=\left [ \tilde{\tilde{n}}_{t|M} \right ]_{1 \leq t \leq m_N}$,
 $\tilde{\tilde{\bm{n}}}_{U}=\left [ \tilde{\tilde{n}}_{t|U}\right ]_{1 \leq t \leq m_N}$,
 $\tilde{\tilde{\bm{n}}}=\left [ \tilde{\tilde{n}}_{t}\right ]_{1 \leq t \leq m_N}=
 \tilde{\tilde{\bm{n}}}_M+\tilde{\tilde{\bm{n}}}_U$,
 $\tilde{\tilde{\bm{p}}}_N =
  \left [ \tilde{\tilde{p}}_{Nt} \right ]_{1 \leq t \leq m_N}$, and
 $\tilde{\tilde{\bm{\lambda}}}_N =
 \left [ \tilde{\tilde{\lambda}}_{Nt} \right ]_{1 \leq t \leq m_N}$,
 such that
 $$\left ( \tilde{\tilde{\bm{p}}}_N,
   (N-1) \tilde{\tilde{\bm{\lambda}}}_N \right ) \stackrel{d}{=}
   \left ( \bm{p}_N, (N-1) \bm{\lambda}_N\right ),$$
 $\tilde{\tilde{n}}_{1|M},\ldots,\tilde{\tilde{n}}_{m_N|M}$,
 $\tilde{\tilde{n}}_{1|U},\ldots,\tilde{\tilde{n}}_{m_N|U}$ are
 conditionally independent given
 $\tilde{\tilde{\bm{p}}}_N$ and $\tilde{\tilde{\bm{\lambda}}}_N$ and
%
%
 \begin{eqnarray*}
 \tilde{\tilde{n}}_{t|M} \left |
 \left (\tilde{\tilde{\bm{p}}}_N, \tilde{\tilde{\bm{\lambda}}}_N \right ) \right .
 &\sim& \mbox{Bernoulli} \left ( \tilde{\tilde{p}}_{Nt} \right ) \\[6pt] 
 \tilde{\tilde{n}}_{t|U}  \left |
 \left (\tilde{\tilde{\bm{p}}}_N, \tilde{\tilde{\bm{\lambda}}}_N \right ) \right .
 &\sim& \mbox{Poisson} \left ( (N-1) \tilde{\tilde{\lambda}}_{Nt} \right )
 \end{eqnarray*}
 Then for any $\bm{T}_N: I\!\!N^{m_N} \rightarrow I\!\!R^d$ and
 $\bm{\omega} \in I\!\!R^d$
%
%
\begin{equation}
 \left |
 E \left [ e^{\jmath \bm{\omega}^{\top} \bm{T}_N \left ( \bm{n} \right ) } \right ] -
 E \left [ e^{\jmath \bm{\omega}^{\top} \bm{T}_N \left ( \tilde{\tilde{\bm{n}}} \right )}
 \right ] \right | =
 O \left ( m_N^2/N\right ),
\end{equation}
 where $\jmath^2=-1$.

\end{theorem}

\noindent \underline{Proof}:
 This result follows from Lemma~(\ref{lemma: first bound characteristic function}) and Lemma~(\ref{lemma: second bound characteristic function}).

\noindent We have the following important corollary.
%
%
\begin{corollary} \label{corollary theorem asymptotic independence}

\noindent Let $\tilde{\tilde{\bm{n}}}=
\left [ \tilde{\tilde{n}}_t \right ]_{1 \leq t \leq m_N}$ and $\bm{T}_N$
 be as in Theorem~\ref{theorem asymptotic independence} and such that
$\bm{T}_N \left ( \tilde{\tilde{\bm{n}}} \right ) \stackrel{d}{\rightarrow} \bm{\xi}$. Then
$\bm{T}_N \left ( \bm{n} \right ) \stackrel{d}{\rightarrow} \bm{\xi}$.

\end{corollary}


\noindent {\it Law of Large Numbers (LLN)}:
 A special case of Corollary~\ref{corollary theorem asymptotic independence} 
 is when 
 {\small $\bm{T}_N \left (\bm{n}\right )=
  m_N^{-1} \sum_{i=1}^{m_N} \bm{T} \left ( n_i \right )$}, where 
 {\small $\bm{T}_N \left ( \tilde{\tilde{\bm{n}}} \right ) \stackrel{p}{\rightarrow}
  E \left [ \bm{T} \left ( \tilde{\tilde{{n}}}_1 \right ) \right ]$} and thus
 {\small $\bm{T}_N \left ( \tilde{\tilde{\bm{n}}} \right ) \stackrel{d}{\rightarrow}
  E \left [ \bm{T} \left ( \tilde{\tilde{{n}}}_1 \right ) \right ]$}.
 According to the corollary, 
 we have  {\small $\bm{T}_N \left (\bm{n}\right ) \stackrel{d}{\rightarrow}
  E \left [ \bm{T} \left ( \tilde{\tilde{{n}}}_1 \right ) \right ]$}, which implies
  {\small $\bm{T}_N \left (\bm{n}\right ) \stackrel{p}{\rightarrow}
  E \left [ \bm{T} \left ( \tilde{\tilde{{n}}}_1 \right ) \right ]$}.

\vspace{6pt}


\noindent {\it Central Limit Theorem (CLT)}: Another special case is when 
 {\small $\bm{T}_N \left (\bm{n}\right )=
   m_N^{-1/2} \sum_{i=1}^{m_N} \bm{T} \left ( n_i \right )$},
 where $E \left [ \bm{T} \left ( \tilde{\tilde{{n}}}_1 \right ) \right ]=\bm{0}$
 and
 $E \left [ \left \| \bm{T} \left ( \tilde{\tilde{{n}}}_1 \right ) \right \|^2\right ]<\infty$
 (where $\| .\|$ is the euclidian norm).
 Then, according to \cite[Theorem 29.5]{billingsley_1995}, we have
 {\small $\bm{T}_N \left ( \tilde{\tilde{\bm{n}}} \right ) \stackrel{d}{\rightarrow}
   N \left (\bm{0},\right )$} and
 {\small $\bm{T}_N \left (\bm{n}\right ) \stackrel{d}{\rightarrow}
 N \left (\bm{0}, \bm{\Sigma}\right )$},
 according to Corollary~\ref{corollary theorem asymptotic independence},
 where $\bm{\Sigma} = E \left [ \bm{T} \left ( \tilde{\tilde{{n}}}_1 \right )
     \bm{T} \left ( \tilde{\tilde{{n}}}_1 \right )^{\top}  \right ]$.

%
\noindent {\it Optimal sample size}: When the above conditions are met, we have
%
%
\begin{equation}
 E \left [ e^{\jmath \bm{\omega}^{\top}
 \bm{T}_N \left ( \tilde{\tilde{\bm{n}}}  \right )} \right ] =
 e^{\bm{\omega}^{\top} \bm{\Sigma} \bm{\omega}/2} + O \left ( m_N^{-1/2}\right ).
\end{equation}
 Hence
%
%
\begin{equation}
 \left |
 E \left [ e^{\jmath \bm{\omega}^{\top} \bm{T}_N \left (\bm{n}\right )} \right ] -
 e^{\bm{\omega}^{\top} \bm{\Sigma} \bm{\omega}/2}  \right |=
 O \left ( m_N^2/N + m_N^{-1/2}\right )
\end{equation}
 The right-hand side of the above equation is minimized when $m_N^2/N = m_N^{-1/2}$, i.e. when
 $m_N=O \left ( N^{2/5}\right )$; the optimal order of magnitude for the sample size.

\vspace{6pt}


\noindent {\it Consistency and asymptotic normality of $\widehat{\bm{\psi}}_N$}:
We can apply Corollary~\ref{corollary theorem asymptotic independence} to examine the consistency and asymptotic normality of $\widehat{\bm{\psi}}_N$, which we view as the function $\bm{\psi}_N(.)$ of $n_1,\ldots,n_{m_N}$.
 Let
 $\tilde{\tilde{\bm{\psi}}}_N =
 \bm{\psi}_N \left ( \tilde{\tilde{n}}_1,\ldots,
 \tilde{\tilde{n}}_{m_N} \right )$.
 If
 $\tilde{\tilde{{\bm{\psi}}}}_N$ is consistent and
 asymptotic normal, i.e.
 $\tilde{\tilde{{\bm{\psi}}}}_N \stackrel{p}{\rightarrow} \bm{\psi}_0$ and
 $m_N^{1/2}\left ( \tilde{\tilde{\bm{\psi}}}_N -
 \bm{\psi}_0 \right ) \stackrel{d}{\rightarrow}
 N \left ( \bm{0}, \bm{\Sigma}\right )$,
 according to Corollary~\ref{corollary theorem asymptotic independence},
 we also have $\widehat{\bm{\psi}}_N \stackrel{p}{\rightarrow} \bm{\psi}_0$ and
 $m_N^{1/2} \left ( \widehat{\bm{\psi}}_N - \bm{\psi}_0 \right ) \stackrel{d}{\rightarrow}
 N \left ( \bm{0}, \bm{\Sigma}\right )$, i.e. $\widehat{\bm{\psi}}_N$ is also consistent and asymptotically normal,
where $\bm{\psi}_0$ is the true value of the parameter vector.
Van der Vaart \cite{vandervaart_1998} has given sufficient conditions for the consistency and
 asymptotic normality of the MLE $\tilde{\tilde{\bm{\psi}}}_N$.
 It is consistent if the parameter $\bm{\psi}$ is identifiable and the log-likelihood
 $\ell \left ( \bm{\psi};k \right ) =
 \log P \left ( \tilde{\tilde{n}}_i=k; \bm{\psi} \right )$ satisfies the following conditions \cite[chap. 5.2, chap. 5.5]{vandervaart_1998}.
\begin{itemize}
\item[C1.] The parameter $\bm{\psi}$ belongs to a compact subset $\Theta$.
\item[C2.] The log-likelihood is a continuous function of $\bm{\psi}$ for every $k$.
\item[C3.] The log-likelihood is dominated by an integrable function, i.e.
 $\left | \ell \left ( \bm{\psi};k \right )  \right | \leq h(k)$ for any $k$,
 where $E \left [ h \left ( \tilde{\tilde{n}}_i \right ) \right ]=
 \sum_{k=0}^{\infty} P \left ( \tilde{\tilde{n}}_i=k; \bm{\psi}_0 \right )
 h \left ( k \right ) <\infty$.
\end{itemize}
The MLE $\tilde{\tilde{\bm{\psi}}}_N$ is asymptotically normal if the following conditions are met \cite[chap. 5.5, Theorem 5.39]{vandervaart_1998}.
\begin{itemize}
\item[N1.] The true parameter value $\bm{\psi}_0$ is in the interior of $\Theta$.
\item[N2.] The MLE $\tilde{\tilde{\bm{\psi}}}_N$ is consistent.
\item[N3.] There exists a measurable function $c: I\!\!N \mapsto \left ( 0, \infty\right )$
 such that
\begin{itemize}
\item[N3.a] The function $c(.)$ is square integrable at $\bm{\psi}_0$, i.e.
 {\small $E \left [ c \left ( \tilde{\tilde{n}}_i \right )^2 \right ]<\infty$}.
\item[N3.b] There exists some neighbourhood\footnote{Here a neighbourhood is an open subset
 containing $\bm{\psi}_0$, e.g., an open ball centered at $\bm{\psi}_0$.} of $\bm{\psi}_0$ where
$ \left | \ell \left ( \bm{\psi}_1; k \right ) -
 \ell \left ( \bm{\psi}_2; k \right )  \right | \leq
 c(k) \left \| \bm{\psi}_1 - \bm{\psi}_2 \right \|$ for any $\bm{\psi}_1$ and $\bm{\psi}_2$
 therein and for any $k$.
\end{itemize}

\item[N4.] There exists a measurable function
 $\dot{\bm{\ell}}: I\!\!N \mapsto I\!\!R^{3G-1}$, such that
%
%
{\small
\begin{eqnarray*}
\sum_{k=0}^{\infty}
\left ( \sqrt{P \left ( \tilde{\tilde{n}}_i=k; \bm{\psi} \right )} -
\sqrt{P \left ( \tilde{\tilde{n}}_i=k; \bm{\psi}_0 \right )} -
\frac{1}{2}\sqrt{P \left ( \tilde{\tilde{n}}_i=k; \bm{\psi}_0 \right )}
\left ( \bm{\psi} - \bm{\psi}_0 \right )^{\top} \dot{\bm{\ell}} \left ( k \right ) \right )^2
&=& \nonumber \\
 o \left ( \left \| \bm{\psi} -\bm{\psi}_0 \right \|^2\right ), & &
\end{eqnarray*}}
 as $\bm{\psi} \rightarrow \bm{\psi}_0$.
 In other words, the log-likelihood is {\it differentiable in the quadratic mean}.

\item[N5.] The log-likelihood is twice differentiable at $\bm{\psi}_0$
 with a nonsingular Fisher information matrix 
 $\bm{I} \left ( \bm{\psi}_0\right ) =
 E \left [ - \left . \left ( \partial^2/ \partial \bm{\psi} \partial \bm{\psi}^{\top} \right )
 \ell \left ( \bm{\psi}; \tilde{\tilde{n}}_i\right ) \right |_{\bm{\psi}=\bm{\psi}_0} \right ]$.
\end{itemize}
 When all these conditions are met,
 $m_N^{1/2}\left ( \tilde{\tilde{\bm{\psi}}}_N -
 \bm{\psi}_0 \right ) \stackrel{d}{\rightarrow}
 N \left ( \bm{0}, \bm{I} \left ( \bm{\psi}_0\right )^{-1} \right )$, and
 $m_N^{1/2} \left ( \widehat{\bm{\psi}}_N - \bm{\psi}_0 \right ) \stackrel{d}{\rightarrow}
 N \left ( \bm{0}, \bm{I} \left ( \bm{\psi}_0\right )^{-1} \right )$.


\noindent {\it Likelihood Ratio Test} (LRT): The LRT may be used to test a null hypothesis $H_0$ against the alternative $H_1$.
 The corresponding statistic converges in distribution to a chi-squared distribution
 under mild conditions that essentially correspond to conditions N.1-N.5 for the unrestircted MLE
 $\tilde{\tilde{\bm{\psi}}}_N$ and the MLE $\tilde{\tilde{\bm{\psi}}}_{N0}$ 
 under the null hypothesis $H_0$ \cite[Theorem 16.7]{vandervaart_1998}.
 By Corollary~\ref{corollary theorem asymptotic independence}, we also have the same asymptotic
 chi-squared distribution for the corresponding statistic based on $n_1,\ldots,n_{m_N}$.
 However this limit does not apply for statistical inference about the number of
 mixture components; an application that is further discussed below.


\subsection{Point estimator}
\label{subsection: point estimator}

\noindent For the estimation, consider that
 $\left ( p_N \left ( \bm{v}_i\right ),  (N-1) \lambda_N \left ( \bm{v}_i\right ) \right )$
 is distributed according to the joint CDF
%
%
 \begin{equation}
 F(p,\lambda) = \sum_{g=1}^G \alpha_{(g)} I
 \left ( p_{(g)} \leq p \ and \ \lambda_{(g)} \leq \lambda\right ),
 \end{equation}
 where $\lambda_{(1)}>\ldots>\lambda_{(G)}$ and $\min_g \alpha_{(g)}>0$.
 Also consider that $n_i$ is approximately distributed according to a finite mixture with
 $G$ components where the $g$-th component is distributed according to
 $Bernoulli \left ( p_{(g)} \right ) + Poisson (\lambda_{(g)})$.

\noindent Let
%
%
\begin{eqnarray*}
n_{i|M} &=&
 I \left ( \bm{v}_{\pi(i)}' \in {\cal B}_N \left ( \bm{v}_i \right ) \right ), \\
n_{i|U} &=&
\sum_{i'\neq i} I \left ( \bm{v}_{\pi(i')}' \in {\cal B}_N \left ( \bm{v}_i \right ) \right ), \\
n_i &=& n_{i|M}+n_{i|U} \\
&=& \sum_{j=1}^N I \left ( \bm{v}_j' \in {\cal B}_N \left ( \bm{v}_i \right ) \right ).
\end{eqnarray*}
 and denote by $\bm{\psi} =
 \left [ \left ( \alpha_{(g)},p_{(g)},\lambda_{(g)} \right ) \right ]_{1 \leq g \leq G}$ the
 vector of parameters.
The estimation is based on the sample $\left [ n_i \right ]_{1 \leq i \leq m_N}$ where
 $m_N=o \left ( N^{-1/2}\right )$ is such that $n_1,\ldots,n_{m_N}$ are nearly independent.
 Let $\widehat{\bm{\psi}}_N$ denote the
 corresponding maximum composite\footnote{The $n_i$'s are correlated.} likelihood estimate. 
 The complete data comprise of
 $\left [ \left ( n_{i|M},n_{i|U},\bm{c}_i \right ) \right ]_{1 \leq i \leq m_N}$, and the
  composite log-likelihood is
%
%
\begin{eqnarray*}
\log L_c &=& \sum_{i=1}^{m_N} \log P(n_{i|M},n_{i|U},\bm{c}_i) \\
&=&
\sum_{i=1}^{m_N} \Bigg (
\sum_{g=1}^G c_{ig} \left ( n_{i|M} \log p_{(g)} + \left ( 1-n_{i|M}\right ) \log \left ( 1 - p_{(g)} \right ) \right ) + \\
& &
\sum_{g=1}^G c_{ig} \left ( n_{i|U} \log \lambda_{(g)} - \lambda_{(g)} -
 \log \left ( n_{i|U}! \right ) \right ) + \sum_{g=1}^G c_{ig} \log \alpha_{(g)} \Bigg )
\end{eqnarray*}
The corresponding maximum likelihood equations are
%
%
\begin{eqnarray*}
\left . \frac{\partial}{\partial p_{(g)}}
\left ( \log L_c - \mu \left ( \sum_{g=1}^G p_{(g)} - 1 \right ) \right ) \right |_{\widehat{p}_{(g)}}&=&
\sum_{i=1}^{m_N}
c_{ig} \left ( \frac{n_{i|M}}{\widehat{p}_{(g)}} -
\frac{1-n_{i|M}}{1-\widehat{p}_{(g)}} \right ) \nonumber \\
&=& 0 \\
%
%
\left . \frac{\partial}{\partial \lambda_{(g)}}
\left ( \log L_c - \mu \left ( \sum_{g=1}^G p_{(g)} - 1 \right ) \right )  \right |_{\widehat{\lambda}_{(g)}}&=&
\sum_{i=1}^{m_N} c_{ig} \left ( \frac{n_{i|U}}{\widehat{\lambda}_{(g)}} - 1 \right ) \\
&=& 0 \\
%
%
\left . \frac{\partial}{\partial \alpha_{(g)}}
\left ( \log L_c - \mu \left ( \sum_{g=1}^G p_{(g)} - 1 \right ) \right )  \right |_{\widehat{\alpha}_{(g)}}&=&
\sum_{i=1}^{m_N} \left ( \frac{c_{ig}}{\widehat{\alpha}_{(g)}} - \mu \right ) \\
&=& 0
\end{eqnarray*}
 where $\mu$ is a Lagrange multiplier. Hence
%
%
\begin{eqnarray}
\widehat{p}_{(g)} &=&
 \frac{\sum_{i=1}^{m_N} c_{ig} n_{i|M}}{\sum_{i=1}^{m_N} c_{ig}} \\[6pt]
%
%
\widehat{\lambda}_{(g)} &=&
 \frac{\sum_{i=1}^{m_N} c_{ig} n_{i|U}}{\sum_{i=1}^{m_N} c_{ig}} \\[6pt]
%
%
\widehat{\alpha}_{(g)} &=& \frac{1}{m_N}\sum_{i=1}^{m_N} c_{ig}
\end{eqnarray}

%
\noindent The observed data comprise of $\left [ n_i \right ]_{1 \leq i \leq m_N}$ and the corresponding composite log-likelihood is
%
%
\begin{eqnarray}
\log L &=&
\sum_{i=1}^{m_N} \log \Bigg ( \sum_{g=1}^G \alpha_{(g)} \Bigg (  I(n_i=0) (1-p_{(g)}) e^{-\lambda_{(g)}} + \nonumber \\
 & &
I(n_i>0) \left ( p_{(g)} + (1-p_{(g)}) \frac{\lambda_{(g)}}{n_i}\right ) \frac{e^{-\lambda_{(g)}} \lambda_{(g)}^{n_i-1}}{(n_i-1)!} \Bigg ) \Bigg )
\end{eqnarray}
%
%
\noindent For class $g$, the corresponding estimates are
\begin{eqnarray}
\label{eq: M-step 1}
\widehat{p}_{(g)} &=&
 \frac{\sum_{i=1}^{m_N} E \left [ c_{ig} n_{i|M} \big | n_i; \widehat{\bm{\psi}}_N \right ]}{\sum_{i=1}^{m_N} E \left [ c_{ig} \big | n_i; \widehat{\bm{\psi}}_N \right ]} \\[6pt]
%
%
\label{eq: M-step 2}
\widehat{\lambda}_{(g)} &=&
 \frac{\sum_{i=1}^{m_N} E \left [ c_{ig} n_{i|U} \big | n_i; \widehat{\bm{\psi}}_N \right ]}{\sum_{i=1}^{m_N} E \left [ c_{ig} \big | n_i; \widehat{\bm{\psi}}_N \right ]} \\[6pt]
%
%
\label{eq: M-step 3}
\widehat{\alpha}_{(g)} &=&
 \frac{1}{m_N}\sum_{i=1}^{m_N} E \left [ c_{ig} \big | n_i; \widehat{\bm{\psi}}_N \right ]
\end{eqnarray}
 where the conditional expectations are computed in the E-step.

%
\noindent The EM algorithm alternates between the E step (based on Eqs. (\ref{eq: E-step 1}), (\ref{eq: E-step 2}),
 and (\ref{eq: E-step 3} in the supplementary material), and Eq.~(\ref{eq: E-step 2})) and the M step (based on Eqs. (\ref{eq: M-step 1}
), (\ref{eq: M-step 2}) and (\ref{eq: M-step 3})).


\subsection{Variance and confidence interval}

\noindent The variance of the point estimator $\widehat{\bm{\psi}}_N$ may be
estimated by the bootstrap method, which consists in drawing many independent
samples with replacement from $n_1,\ldots,n_{m_N}$ and computing an estimate
for each such sample.
The bootstrap variance is the sample variance across these different estimates.



\noindent For each parameter of interest, we compute a $100(1-\alpha)\%$ normal confidence
 interval using the estimated variance.


\subsection{Number of components}

\noindent Many authors have studied the difficult problem of making valid inferences about the number of mixture components, including MacLachlan \cite{maclachlan_1987},
Feng and McCulloch \cite{feng_mcculloch_1996},
Lo et al. \cite{lo_mendell_rubin_2001}, Garel and Goussanou \cite{garel_goussanou_2002} and 
Li et al. \cite{chen_li_fu_2012}, to name just a few.
 In all these previous studies, the solution relies on a Likelihood Ratio Test (LRT).
 However the related statistic does not have the usual chi-squared distribution,
 because the null hypothesis places the parameter on the boundary of its space.
So far, the proposed methods have included the analytical derivation of the correct limiting distribution \cite{lo_mendell_rubin_2001,garel_goussanou_2002}, the modification of the likelihood with a penalty \cite{chen_li_fu_2012} and the use of a parametric bootstrap \cite{maclachlan_1987,feng_mcculloch_1996}.

\noindent In this work, we apply the parametric bootstrap technique that is described by Feng and McCulloch \cite{feng_mcculloch_1996}.
The validity of the procedure does not require the identification property.
However, further assumptions are made beyond conditions N1-N5, including the existence and
 integrability of third order partial derivatives.
For the problem at hand, this procedure operates on the premise that the mixture has no more than $G$ components, for some known and fixed $G$.
It tests the null hypothesis that the mixture has $G_0<G$ components against the alternative that it has the maximum number of components. 
The procedure has three steps. 
In the first step, the MLE $\widehat{\bm{\psi}}_N$ is computed based on the null hypothesis.
In the second step, the critical level of the LRT is estimated by drawing independent iid samples (according to null distribution and the MLE) and by computing the LRT statistic for each such sample. 
Then, the critical level is set to the proper sample quantile for the resulting LRT statistics. 
In the third step, the LRT is performed using the original observations and the estimated critical level.


\section{Simulations}\label{section: simulations}

\noindent For the simulations, we consider two registers of the same population
 with $N=32,000$ individuals.
 The records comprise of $K=15$ binary variables, i.e. $\bm{v}_i$ and
 $\bm{v}_j'$ are from $\{0,1\}^K$.
 The linkage variables and errors are generated according to the
 following model.
 For $i=1,\ldots,N$
%
%
\begin{eqnarray*}
 u_1,\ldots,u_N &\stackrel{iid}{\sim}& N(0,\sigma_u^2),\\[3pt]
 \left . v_{i1},\ldots,v_{iK} \right |u_i &\stackrel{iid}{\sim}& {Bernoulli} (\mu_i),\\[3pt]
 \mu_i &=& \left ( 1 + e^{-\beta_0-\beta_1 u_i} \right )^{-1},\\[3pt]
 u_1',\ldots,u_N' &\stackrel{iid}{\sim}& N(0,\sigma_{u'}^2),
\end{eqnarray*}
$v_{\pi(i)1}',\ldots,v_{\pi(i)K}'$ are independent given $u_i'$ and
$\left [ v_{ik} \right ]_{1 \leq k \leq K}$ such that
\begin{equation*}
 P \left ( \left . v_{\pi(i)k}'\neq v_{ik} \right |
 u_i', \left [ v_{ik} \right ]_{1 \leq k \leq K}\right ) =
 \mu_i'= \left ( 1 + e^{-\beta_0'-\beta_1' u_i'} \right )^{-1}.
\end{equation*}
 A neighbour is defined as record where all the variables agree, i.e. 
 ${\cal B}_N \left ( \bm{v}_i \right )=  \left \{ \bm{v}_i \right \}$.
 The parameters of interest are $E[p] = P \left ( \left . \bm{v}_j' \in
 {\cal B}_N \left ( \bm{v}_i\right ) \right | M \right )$ and
 $E[\lambda] = (N-1) P \left ( \left . \bm{v}_j' \in
 {\cal B}_N \left ( \bm{v}_i\right ) \right | U \right )$.
 We consider two scenarios.
 In the first scenario, the linkage variables and  
 recording errors are mutually independent across
 the different variables, based on
 the setting $\beta_0=0.0,\beta_1=0.0, \beta_0'=-5.0$ and
 $\beta_1'=0.0$.
 In the second scenario, the linkage variables and
 recording errors are dependent across the different variables,
 based on the setting $\beta_0=0.0, \beta_1=0.5, \beta_0'=-5.0$,
 and $\beta_1'=-0.5$.
 For each scenario, we use 100 repetitions and $m_N=2 \left \lceil N^{2/5} \right \rceil = 128$
 (note that this is compatible with the requirement $m_N = o(\sqrt{N})$ for the asymptotic independence of the $n_i$'s).
 With the proposed model, the parameters are estimated using a simple random sample of $m_N$
 observations, where each observation is the number of neighbours of some record in the first file.
 The number of classes $G$ is varied between 1 and 5.
The other model assumes the conditional independence of the linkage variables.
For this model, the estimation of the parameters $E[p]$ and $E[\lambda]$ is based on the sample that is comprised of all 32,000$\times$32,000=1.024B pairs in the Cartesian product of the two files\footnote{For computational efficiency, the E-M algorithm only uses the frequency dataset with  no more than $2^K=32,768$ rows, where each row gives the number of observed pairs for some
 $\gamma \in \{0,1\}^K$.}.
For the pair $(i,j)$, the observation is the comparison vector
 $\gamma_{ij}=\left [ \gamma_{ij}^{(k)}\right ]_{1\leq k \leq K} \in \{ 0,1\}^K$, where
 $\gamma_{ij}^{(k)}=I \left ( v_{ik}=v_{jk}'\right )$.
Under the conditional independence assumption, the unknown parameters are
 $\left [ \left ( P \left ( \left . \gamma^{(k)}=1 \right | M\right ),
 P \left ( \left . \gamma^{(k)}=1 \right | U\right ) \right )
 \right ]_{1 \leq k \leq K}$, which are related to the parameters of interest by the following equations.
%
%
\begin{eqnarray}
E[p]&=& P \left ( \left . \gamma^{(k)}=1, \ k=1,\ldots,K \right | M \right ) \\ [6pt]
&=& \prod_{k=1}^K P \left ( \left . \gamma^{(k)}=1 \right | M \right ), \\ [6pt]
E[\lambda]&=& P \left ( \left . \gamma^{(k)}=1, \ k=1,\ldots,K \right | U \right ) \\ [6pt]
&=&(N-1)\prod_{k=1}^K P \left ( \left . \gamma^{(k)}=1 \right | U \right ).
\end{eqnarray}
The model parameters are estimated by maximizing the (composite) likelihood of all the record pairs, using the known mixing proportion of $1/N$.
The E-M algorithm is based on the following equations
%
%
{\small
\begin{eqnarray}
\label{eq: classical E-step}
\widehat{P} \left ( M \left | \gamma_{ij} \right .\right ) &=&
\left ( 1 + \left ( N-1 \right )
\frac{ \prod_{k=1}^K \widehat{P} \left ( \left . \gamma^{(k)}=1 \right | U\right )^{\gamma_{ij}^{(k)}} \widehat{P} \left ( \left . \gamma^{(k)}=0 \right | U\right )^{1-\gamma_{ij}^{(k)}} }{
\prod_{k=1}^K \widehat{P} \left ( \left . \gamma^{(k)}=1 \right | M\right )^{\gamma_{ij}^{(k)}} \widehat{P} \left ( \left . \gamma^{(k)}=0 \right | M\right )^{1-\gamma_{ij}^{(k)}} }
\right )^{-1},\\[6pt]
%
%
\label{eq: classical M-step matched}
\widehat{P} \left ( \left . \gamma^{(k)}=1 \right | M\right ) &=&
\frac{\sum_{i=1}^N \sum_{j=1}^N \widehat{P} \left ( M \left | \gamma_{ij} \right .\right ) 
I \left ( \gamma_{ij}^{(k)}=1\right )}{
\sum_{i=1}^N \sum_{j=1}^N \widehat{P} \left ( M \left | \gamma_{ij} \right .\right )}, \\[6pt]
%
%
\label{eq: classical M-step unmatched}
\widehat{P} \left ( \left . \gamma^{(k)}=1 \right | U\right ) &=&
\frac{\sum_{i=1}^N \sum_{j=1}^N \widehat{P} \left ( U \left | \gamma_{ij} \right .\right ) 
I \left ( \gamma_{ij}^{(k)}=1\right )}{
\sum_{i=1}^N \sum_{j=1}^N \widehat{P} \left ( U \left | \gamma_{ij} \right .\right )},
\end{eqnarray}}
 where the E-step is based on Eq.~(\ref{eq: classical E-step}) and the M-step is based on
 Eq.~(\ref{eq: classical M-step matched}) and Eq.~(\ref{eq: classical M-step unmatched}).

\noindent The results for the first simulation scenario are given in
 Tables~\ref{table: accuracy in the first simulation scenario} and
 \ref{table: QQ plots for the first simulation scenario}.
Table~\ref{table: accuracy in the first simulation scenario} shows that both models estimate the parameters $E[p]$ and $E[\lambda]$ with a small relative bias and a small mean squared error.
In most cases, the achieved performance is better with the estimator that is based on the conditional independence assumption as seen from the smaller relative bias, and from the standard error and mean squared error that are smaller by one or two orders of magnitude.
For the proposed estimator, the achieved mean squared error varies little when the number of classes is increased from 1 to 5 and it is smallest when using $G=4$ classes; the optimal number of classes.
In Table~\ref{table: QQ plots for the first simulation scenario}, the QQ plots show that for the proposed model, the estimators have an approximately normal distribution, in agreement with the discussion in Section~\ref{subsection: Asymptotic independence}.
Overall, the above results demonstrate a superior performance by the model based on the conditional independence assumption.
This may be because it accounts for the structure of correlation among the linkage variables or because of the far greater sample size, i.e. 1.024B pairs vs. 128 $n_i$'s.
They also show that when using this model, the mixing proportion can be significantly smaller than 5\%, provided it is a known parameter and conditional independence applies.
Indeed, among record linkage practitioners it was believed that unsurmountable convergence issues would arise with the related E-M algorithm if this condition were not satisfied.
The results from the first scenario clearly demonstrate the need to nuance this opinion, since the mixing proportion is 1/32,000 and smaller than 5\% by three orders of magnitude.
Finally, these results suggest that the proposed estimators may be  improved by accounting for the correlation structure of the linkage variables and by increasing the sample size.

%
{\footnotesize
\begin{table}[htbp]
\begin{center}
\caption{Accuracy in the first simulation scenario.
\label{table: accuracy in the first simulation scenario}}
\begin{tabular}{lcccccccc}\hline
& & \multicolumn{3}{c}{$E[p]$} & & \multicolumn{3}{c}{$E[\lambda]$} \\ \cline{3-5}\cline{7-9}
& & Bias(\%) & SE & MSE &
  & Bias(\%) & SE & MSE \\ \hline 
Model with CI & & 1.73 & 0.0079 & 0.00030 &
& 2.46 & 0.00018 & 0.00064 \\ [-6pt]
New model & $G$ & \\ [-6pt]
 		& 1  & -2.04 & 0.059 & 0.0038 &
 		& 0.10  & 0.11 & 0.012 \\ [-6pt]
		& 2	& 2.27  & 0.050 & 0.0029 &
		& -4.53 & 0.10 & 0.012 \\ [-6pt]
		& 3	& 1.99  & 0.044 & 0.0022 &
		& -5.59 & 0.11 & 0.015 \\ [-6pt]
		& 4	& 3.44  & 0.039 & 0.0024 &
		& -4.00 & 0.10 & 0.011 \\ [-6pt]
		& 5	& 3.28  & 0.042 & 0.0026 &
		& -4.28 & 0.11 & 0.013 \\ \hline
\end{tabular}
\end{center}
\end{table}}

%
{\small
\begin{table}[htbp]
\begin{center}
\caption{QQ plots for the first simulation scenario.
\label{table: QQ plots for the first simulation scenario}}
\begin{tabular}{c}
\includegraphics[width=4.5in]{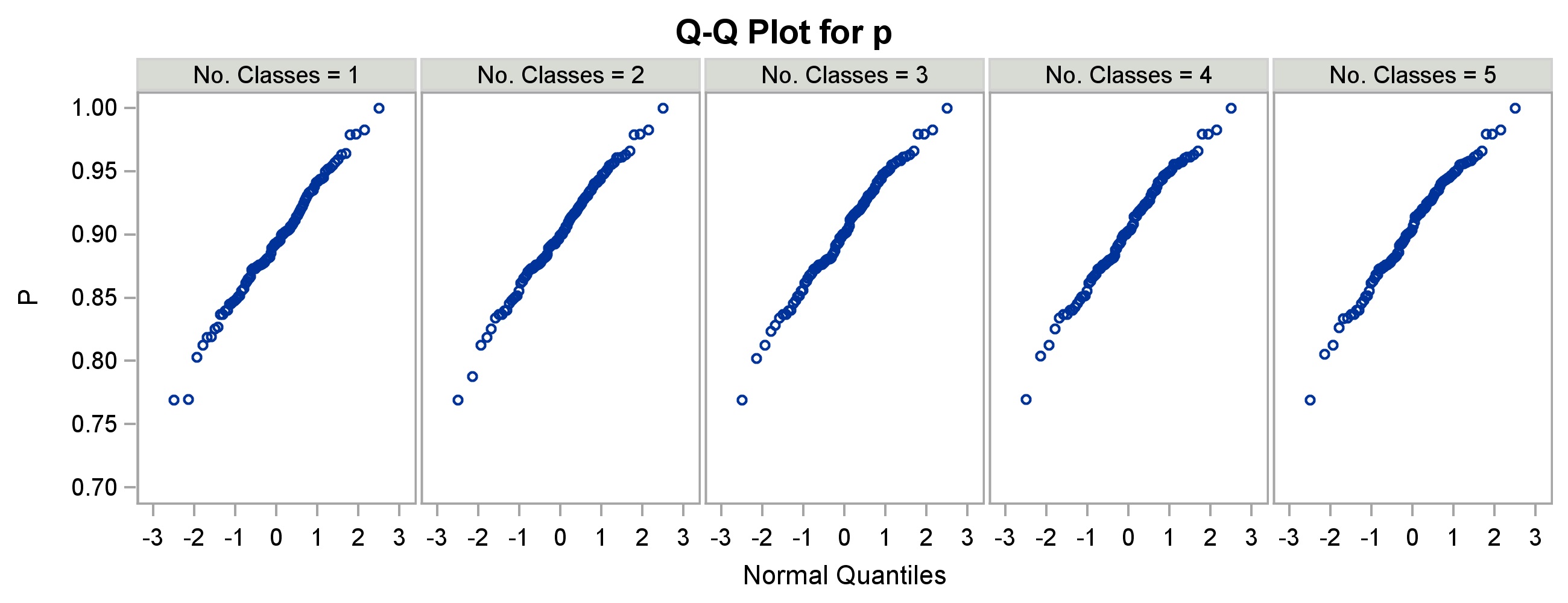} \\
\includegraphics[width=4.5in]{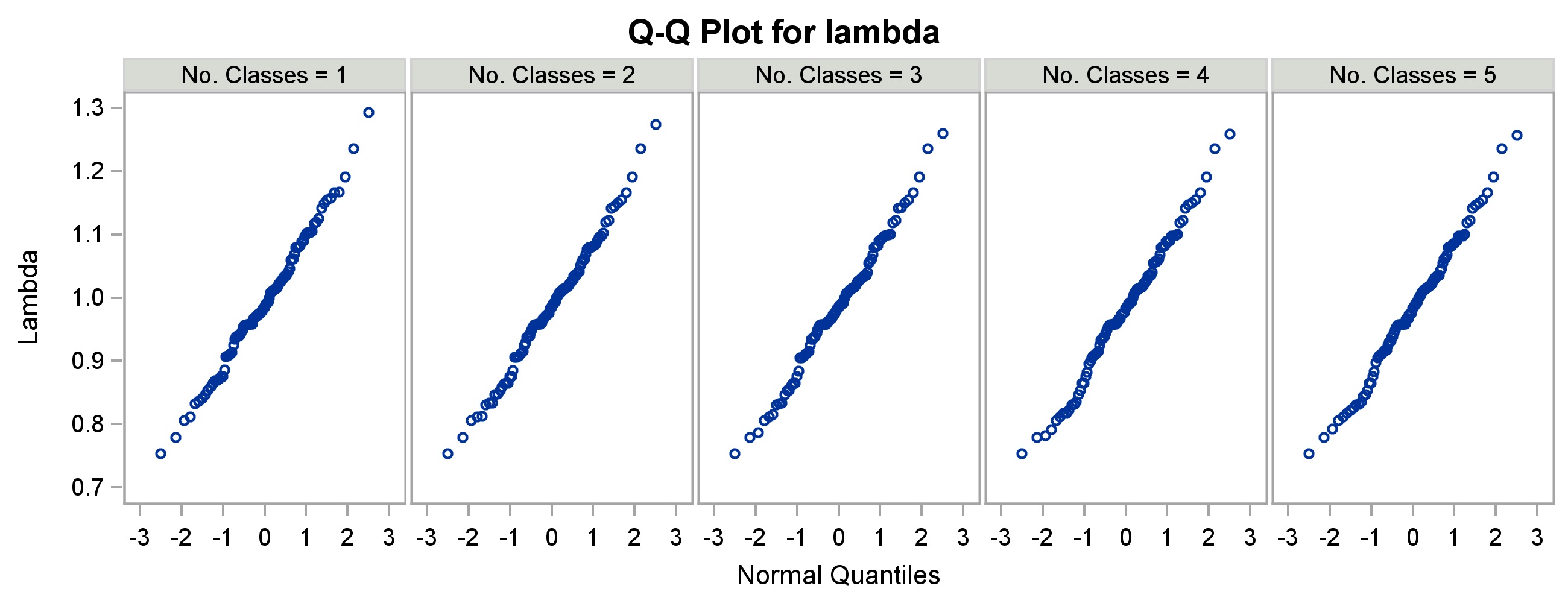}
\end{tabular}
\end{center}
\end{table}}

\noindent The results for the second simulation scenario are given in
 Tables~\ref{table: accuracy in the second simulation scenario} and
 \ref{table: QQ plots for the second simulation scenario}.
 In this scenario, the linkage variables are no longer conditionally independent.
Table~\ref{table: accuracy in the second simulation scenario} shows that only the proposed model estimates both $E[p]$ and $E[\lambda]$ with a small relative bias and a small mean squared error
 if using two or more classes.
 Indeed, under the conditional independence assumption, the estimator of $E[p]$ still has a small relative bias and a small mean squared error.
 However, the estimator of $E[\lambda]$ now has a large relative bias of -27.9\%. 
 Meanwhile, the proposed estimators remain accurate except when using the homogeneous model with $G=1$ classes.
 In this latter case, the estimators have a large relative bias of
 -40.7\% and 23.6\% for $E[p]$ and $E[\lambda]$ respectively.
 This occurs because the homogeneous model does not properly account for the heterogeneity
 of the simulated records.
 When using two or more classes, the proposed estimators have a small relative bias not exceeding 6\% in absolute value.
 In this case, the mean squared error is minimal with $G=2$ and with $G=4$ for $E[p]$ and $E[\lambda]$ respectively.
 Overall, the optimal number of classes may be chosen in between these two values at $G=3$.
In Table~\ref{table: QQ plots for the second simulation scenario}, the QQ plots show that the  proposed estimators are approximately normal distribution when using two or more classes.
Overall, a superior performance is achieved with the proposed estimators.
Also, the above results may suggest that when interested in $E[p]$ only, the classical estimator
 may still be a good choice even if the linkage variables are not conditional independent.
 Yet this proposition must be tested using more extensive simulations.

%
{\footnotesize
\begin{table}[htbp]
\begin{center}
\caption{Accuracy in the second simulation scenario.
\label{table: accuracy in the second simulation scenario}}
\begin{tabular}{lcccccccc}\hline
& & \multicolumn{3}{c}{$E[p]$} & & \multicolumn{3}{c}{$E[\lambda]$} \\ \cline{3-5} \cline{7-9}
& & Bias(\%) & SE & MSE &
  & Bias(\%) & SE & MSE \\ \hline  
Model with CI & & -4.59 & 0.0070 & 0.0017 &
 & -27.9 & 0.00020 & 0.15 \\ [-6pt]
New model  & $G$ & \\ [-6pt]
 		& 1	& -40.7 & 0.41  & 0.30 &
 		& 23.6  & 0.60 & 0.47  \\ [-6pt]
		& 2	& 2.63  & 0.047 & 0.0028 &
		& -3.70 & 0.28 & 0.078 \\ [-6pt]
		& 3	& 4.78  & 0.036 & 0.0031 &
		& -5.06 & 0.26 & 0.070 \\ [-6pt]
		& 4	& 5.42  & 0.032 & 0.0034 &
		& -5.46 & 0.25 & 0.069 \\ [-6pt]
		& 5	& 5.62  & 0.031 & 0.0035 &
		& -5.59 & 0.25 & 0.069 \\ \hline
\end{tabular}
\end{center}
\end{table}}

%
{\small
\begin{table}[htbp]
\begin{center}
\caption{QQ plots for the second simulation scenario.
\label{table: QQ plots for the second simulation scenario}}
\begin{tabular}{c}
\includegraphics[width=4.5in]{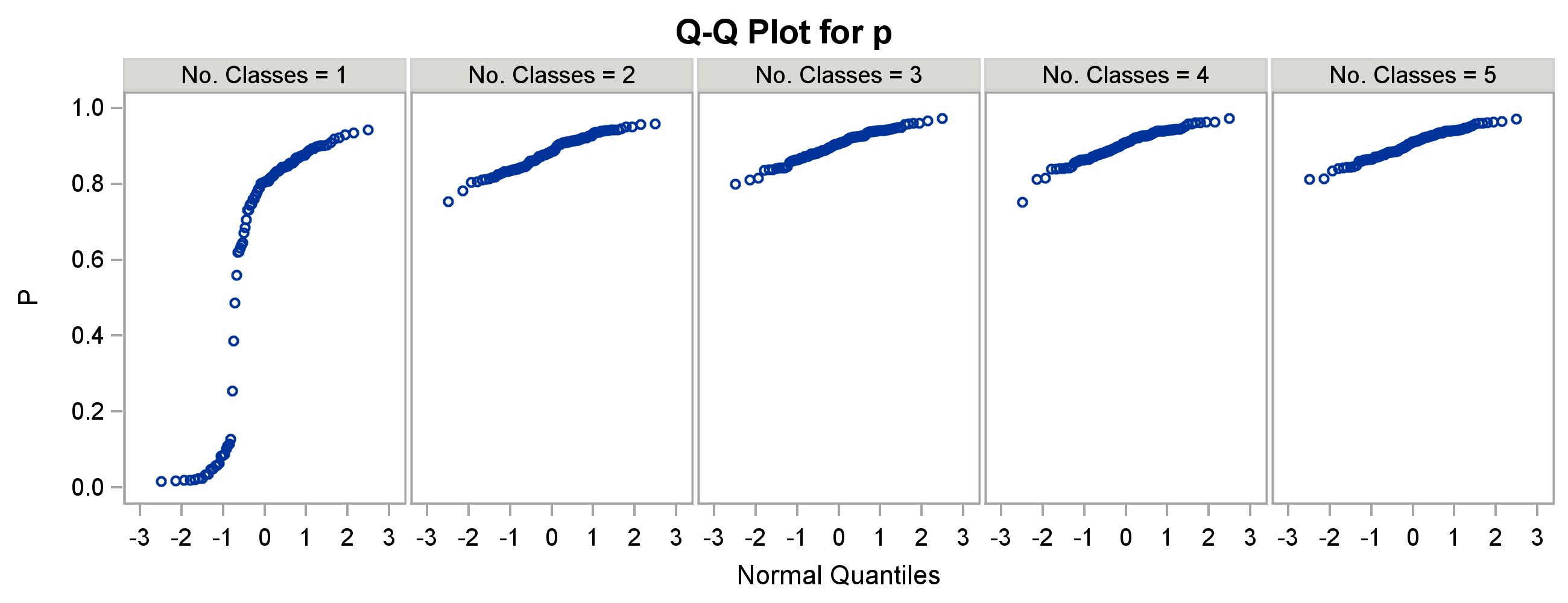} \\
\includegraphics[width=4.5in]{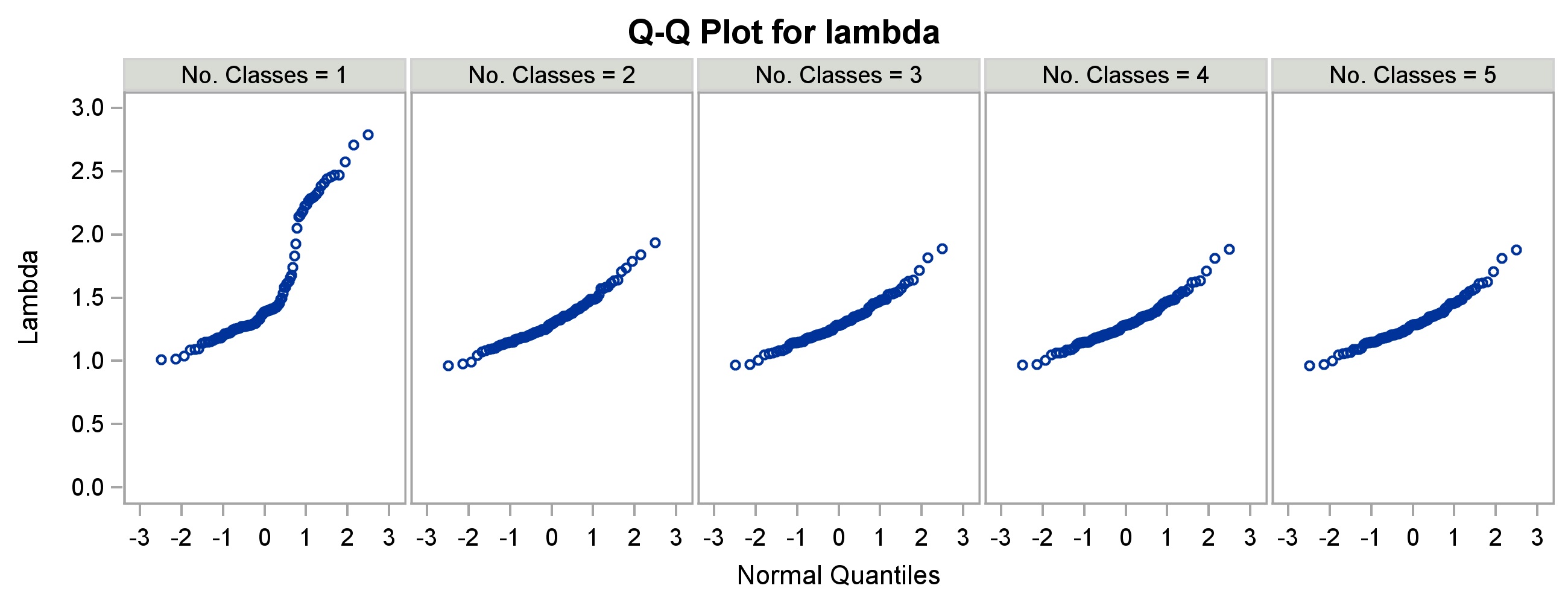}
\end{tabular}
\end{center}
\end{table}}


\section{Empirical study}\label{section: empirical study}

\noindent For the empirical study, we consider the linkage of administrative files representing two years (2012 and 2015) of the same source, using the first family name, the birth date and the city.
 The first file comprises of 25.4M individuals, while the second comprises of 26.9M individuals, including $N=23.9M$ of them in both files.
A unique identifier is available to identify the individuals within and across the two files.
However, It is only used for the evaluation of the proposed estimator.
 For the study, the target population is based on the individuals that are found in both files.
 A blocking key is created by concatenating the year and month of birth, the first three letters
 of the family name SOUNDEX code and the first three letters of the city SOUNDEX code.
 A total of 99.057M potential pairs are selected, including 20.12M true positives and 78.937M false positives as given in the confusion matrix of Table~\ref{table: confusion matrix}.
According to this matrix, the FNR is measured at 15.82\% while the FPR is measured at
0.0000144\%.
The error rates are also estimated with the model using $G=13, 14, 15$ and $16$.
The parameters are estimated using a simple random sample of size $m_N=\left \lceil N^{2/5} \right \rceil = 895$, where each observation is the number of neighbours for a record in the first file.
The E-M algorithm is stopped after 1,000 iterations.
The corresponding parameter estimates are given
 in Table~\ref{table: model estimates for the empirical study}, where the variance and 95\% confidence interval is based on 250 bootstrap samples.
 The point estimate for $E[p]$ varies between 0.8502 and 0.8655, respectively with
 $G=13$ and $G=16$.
 As for $E[\lambda]$ the point estimate varies between 3.3392 and 3.3545, respectively with
 $G=16$ and $G=13$.
 The corresponding error estimates are given in Table~\ref{table: estimated error rates for the empirical study}.
 The actual FNR belongs to the 95\% confidence interval when $G=13,14$ and $15$. The only exception is when $G=16$, with an estimate that is very slightly to the right of the interval.
 As for the actual FPR, it belongs to the 95\% confidence interval for all choices of $G$ between $13$ and $16$.
Table~\ref{table: testing the number of components} shows the results of the likelihood ratio test for $G$ the number of components where the maximum number of components is fixed at 16.
The type I error is set at 0.05. For a given value of $G$, the corresponding critical level
 $c_{\alpha}$ is estimated by a parametric bootstrap procedure as explained previously.
The number of bootstrap samples is 250.
The null hypothesis that there are $G=15$ classes is not rejected at the chosen level.
However, it is rejected for the smaller values of $G$.
These results suggest that we should use 15 classes.

%
{\footnotesize
\begin{table}[htbp]
\begin{center}
\caption{Confusion matrix in millions of pairs.
\label{table: confusion matrix}}
\begin{tabular}{lccc} \hline
	& L		& NL 			& Total	\\	\hline
M	& 20.12	& 3.78		& 23.9	\\[-6pt]
U	& 78.937	& 571.21M		& 571.21M	\\[-6pt]
Total	& 99.057	& 571.21M		& 571.21M	\\ \hline
\end{tabular}
\end{center}
\end{table}}

%
{\footnotesize
\begin{table}[htbp]
\begin{center}
\caption{Model estimates for the empirical study.
\label{table: model estimates for the empirical study}}
\begin{tabular}{cccccccc}\hline
& \multicolumn{3}{c}{$E[p]$} & & \multicolumn{3}{c}{$E[\lambda]$} \\ \cline{2-4} \cline{6-8}
$G$	&	Estimate	&	SE	& 95\% C.I. &
	&	Estimate	&	SE	& 95\% C.I. \\ \hline
13	&	0.8502	&	0.02	& (0.8315,	0.9084) &
	& 3.3545 &	0.26608 & (2.8204, 3.8409)	\\[-6pt]
14	&	0.8596	&	0.01732 & (0.8355, 0.9033) &
	&	3.3451	&	0.2772 & (2.817,	3.9247) \\[-6pt]
15	&	0.8644	&	0.01673 & (0.8381, 0.9046)	&
	&	3.3403	&	0.28169 & (2.8034, 3.9515)	\\[-6pt]
16	&	0.8655	&	0.01517 & (0.8463, 0.9057)	&
	&	3.3392	&	0.25485 & (2.8788, 3.8309)	\\ \hline
\end{tabular}
\end{center}
\end{table}}

%
{\footnotesize
\begin{table}[htbp]
\begin{center}
\caption{Estimated error rates for the empirical study.
\label{table: estimated error rates for the empirical study}}
\begin{tabular}{lccccccc}\hline
& \multicolumn{3}{c}{FNR} & & \multicolumn{3}{c}{FPR} \\ \cline{2-4} \cline{6-8} 
$G$	& Actual & Estimate &	95\% C.I. &
	& Actual & Estimate &	95\% C.I. \\ \hline
13	& 0.1582 & 0.1498 & (0.0916, 0.1685) &
	& 1.4423E-07 & 1.4036E-07 &	(1.18E-07, 1.61E-07)	\\[-6pt]
14	&	 & 0.1404 & (0.0967, 0.1645) &
	& & 1.3996E-07 & (1.18E-07,	1.64E-07) \\[-6pt]
15	&	 & 0.1356 & (0.0954, 0.1619) &
	& &	1.3976E-07	& (1.17E-07, 1.65E-07) \\ [-6pt]
16	&	 & 0.1345 & (0.0943, 0.1537) &
	& &	1.3972E-07	& (1.20E-07, 1.60E-07)	\\ \hline
\end{tabular}
\end{center}
\end{table}}

%
{\footnotesize
\begin{table}[htbp]
\begin{center}
\caption{Testing the number of components.
\label{table: testing the number of components}}
\begin{tabular}{cccc}\hline
$G$ & $\widehat{c}_{\alpha}$ & LR stat. & Result \\	\hline
13	& 0.06081 & 0.08285 &	reject \\[-6pt]
14	& 0.04737 & 0.05362 &	reject \\[-6pt]
15	& 0.01468 & 0.00732 &	accept \\ \hline
\end{tabular}
\end{center}
\end{table}}

\section{Conclusion}
\label{section: future work}

\noindent In this paper, we have addressed the problem of estimating the probabilistic linkage parameters and the rates of linkage error, when linking two population registers that have a perfect coverage, i.e. with no undercoverage and no overcoverage\footnote{Overcoverage comprises of duplicate or erroneous records.}.
 The proposed methodology may be improved in many ways, starting with
 the development of alternative estimators that provide greater
 precision.
 Indeed, convenience is the biggest incentive for the proposed
 methodology, where the selected observations are considered
 independent.
%
%
 However, the resulting estimators only use $o \left ( N^{1/2}\right )$
 observations (the $n_i$'s) and thus have a limited precision not
 exceeding $O \left ( N^{-2/5}\right )$.
 In the future, we will look for alternative estimators that can
 offer a greater precision of $O \left ( N^{-1} \right )$,
 by using $O(N)$ observations; the most obvious candidate being
 the maximum composite likelihood estimator that uses all 
 the $n_i$'s.
 Such estimators will likely differ from
 maximum likelihood estimators regarding their second order properties.
 Although the proposed estimators are of interest mainly because they are oblivious of the correlation structure of the linkage variables, their accuracy may be improved further by accounting for this structure.
Future work must also address more complex linkages where one file is possibly a probability sample and duplicate or incomplete records are found in either file.
 Indeed, the proposed estimators may be biased in
 the presence of duplicate records.
 A simple solution is removing such records from each file
 before the linkage.
 Another solution is keeping them but refining the model
 with additional parameters about the distribution
 of duplicate records.
%
%
 Yet, another alternative may be considered if the original
 files are free of duplicates but duplicate records are created on purpose to facilitate matrix comparisons\footnote{A matrix comparison is the simultaneous comparison of many similar attributes, e.g., first and second last names or day and month in a date, while accounting for potential transpositions or permutations of these attributes.} or to exploit additional information about a given individual, e.g., previous addresses \cite{sanmartin_et_al_2016}.
 In this case, the proposed methodology 
 may be adapted by defining the neighbourhood of
 an original record as the union of neighbourhoods for all the related duplicate records.
 The occurence of records with missing values is another
 practical concern.
 In the simplest scenario, a record is considered
 missing if one variable is missing.
 Even then, there are two distinct issues that vastly differ
 in their impact.
 The first issue occurs when a record $\bm{v}_i$ has a missing value that is causing the related $n_i$ to be missing.
 The second issue is more severe and occurs when some record $\bm{v}_j'$ has a missing value
 that is causing {\it all} the $n_i$'s to be missing.
 To address both issues, we may compute $n_i$ based only on the
 complete $\bm{v}_j'$ records, for each complete $\bm{v}_i$ record.
 Then, we may fit a refined model to the resulting sample; a model
 that may have to account for the missing data mechanism even if the records are
 {\it incomplete completely at random} (ICAR).

\bigskip
\begin{center}
{\large\bf SUPPLEMENTARY MATERIAL}
\end{center}

\begin{description}

\item[Lower bound on the success probability]: Proof of
Eq.~\ref{eq success probability lower bound} in Section~\ref{section: notation and assumptions}.

\item [Lower bound on the Shannon entropy]: Proof of
Eq.~\ref{eq shannon entropy lower bound} in Section~\ref{section: notation and assumptions}.

\item [Linkage parameters and errors]: Derivations of Eqs.~\ref{eq m-probability} to \ref{eq: linkage accuracy at record level} in Section~\ref{section: linkage parameters and errors}, which show the connection between the distribution of the number of neighbours on one hand and the probabilistic linkage parameters and different measures of linkage error on the other hand.

\item[Proof of Theorem 1].

\item[Proof of Theorem 2].

\item[Proof of Corollary 1].

\item[E step]: Details of E-M procedure described in Section~\ref{subsection: point estimator}.

\item[SAS code]: SAS macros used in the empirical study and in  the simulations.

\end{description}


\section{Lower bound on the success probability}

%
\begin{eqnarray}
 \tau &=&
 P \left ( n_{i|M}=1 \right )
 E\left [ n_i^{-1}\left | n_{i|M}=1 \right . \right ]
 \nonumber \\[6pt]
 &=&
 E[ P \left ( \left . \bm{v}_{\pi(i)}' \in {\cal B}_N \left ( \bm{v}_i \right ) \right |
 \bm{v}_i \right )]
 E\left [ \left ( 1+n_{i|U} \right )^{-1}\left | n_{i|M}=1 \right . \right ] \nonumber \\[6pt]
 &\geq&
 \delta \left ( 1 + 
 E\left [ n_{i|U} \left | n_{i|M}=1 \right . \right ]
 \right )^{-1} \nonumber \\[6pt]
 &\geq&
 \frac{\delta}{1+\Lambda},
\end{eqnarray}
 where we have used the convexity of the function $x \mapsto (1+x)^{-1}$ and Jensen's
 inequality \cite[pp. 80]{billingsley_1995}.


\section{Lower bound on the Shannon entropy}

%
\begin{eqnarray}
 H_N &=& E \left [ - \log P \left ( \bm{v}_j'\right )\right ] \nonumber \\
&=&
\sum_{r=1}^{R_N}
P \left ( {\cal B}_N \left ( \bm{v}_{(r)} \right )  \right )
\left ( -\log P \left ( {\cal B}_N \left ( \bm{v}_{(r)} \right )  \right ) \right ) + \nonumber \\
& &
\sum_{r=1}^{R_N}
P \left ( {\cal B}_N \left ( \bm{v}_{(r)} \right )  \right )
E \left [ \left . -\log P \left ( \bm{v}_j' \left | {\cal B}_N \left ( \bm{v}_{(r)} \right )   \right .\right ) \right |
{\cal B}_N \left ( \bm{v}_{(r)} \right )  \right ] \nonumber \\
&=&
\left ( \sum_{r=1}^{R_N} P \left ( {\cal B}_N \left ( \bm{v}_{(r)} \right )
  \right ) \right )
\left ( -\log \left ( \frac{\Lambda}{N-1} \right )  \right )
 + \nonumber \\
& &
\underbrace{\sum_{r=1}^{R_N}
P \left ( {\cal B}_N \left ( \bm{v}_{(r)} \right )  \right )
\left ( -\log \left (
\frac{P \left ( {\cal B}_N \left ( \bm{v}_{(r)} \right )  \right )}{\Lambda/(N-1)} \right ) \right )}_{\geq 0} + \nonumber \\
& &
\underbrace{\sum_{r=1}^{R_N}
P \left ( {\cal B}_N \left ( \bm{v}_{(r)} \right )  \right )
E \left [ \left . -\log P \left ( \bm{v}_j' \left | {\cal B}_N \left ( \bm{v}_{(r)} \right )   \right .\right ) \right |
{\cal B}_N \left ( \bm{v}_{(r)} \right )  \right ]}_{\geq 0} \nonumber \\
&\geq&
\left ( \sum_{r=1}^{R_N}
P \left ( {\cal B}_N \left ( \bm{v}_{(r)} \right )  \right ) \right )
\left ( -\log \left ( \frac{\Lambda}{N-1} \right )  \right ) \nonumber \\
&=&
\left ( 1 - O \left ( N^{-1} \right ) \right )
\left ( -\log \left ( \frac{\Lambda}{N-1} \right )  \right )
\end{eqnarray}


\section{Linkage parameters and errors}

\noindent {\it m-probability}: Using Eq.~(\ref{eq: second condition}) it can be shown that (see the supplementary material)
%
%
\begin{eqnarray*}
P \left ( \left . \bm{v}_j' \in {\cal B}_N \left ( \bm{v}_i \right ) \right | M \right )
 &=&
 P \left ( \left . \bm{v}_j' \in {\cal B}_N \left ( \bm{v}_i \right ) \right |
 j=\pi(i) \right ) \nonumber \\
 &=&
 P \left ( \left . \bm{v}_{\pi(i)}' \in {\cal B}_N \left ( \bm{v}_i \right ) \right |
 j=\pi(i) \right ) \nonumber \\
 &=&
 P \left ( \bm{v}_{\pi(i)}' \in {\cal B}_N \left ( \bm{v}_i \right )\right ) \nonumber \\
 &=&
 E \left [
 P \left ( \left . \bm{v}_{\pi(i)}' \in {\cal B}_N \left ( \bm{v}_i \right )
 \right | \bm{v}_i \right )
 \right ] \nonumber \\
 &=&
 E \left [ p_N \left ( \bm{v}_i \right ) \right ] \nonumber \\
 &=& \int p dF(p,\lambda) \nonumber \\
 &=& E[p].
\end{eqnarray*}

%
\noindent {\it u-probability}: According to Eq.~(\ref{eq: second condition}) we have
%
%
\begin{eqnarray*}
 P \left ( \left . \bm{v}_j' \in {\cal B}_N \left ( \bm{v}_i 
 \right ) \right | U \right )
 &=&
 P \left ( \left . \bm{v}_j' \in
 {\cal B}_N \left ( \bm{v}_i \right ) \right |
 j \neq \pi(i) \right ) \nonumber \\
 &=&
 \sum_{i' \neq i}
 P \left ( \left . j= \pi(i') \right |
 j \neq \pi(i) \right )
 P \left ( \left . \bm{v}_j' \in
 {\cal B}_N \left ( \bm{v}_i \right ) \right |
 j = \pi(i') \right ) \nonumber \\
&=&
 \frac{1}{N-1} \sum_{i' \neq i}
 P \left ( \left . \bm{v}_{\pi(i')}' \in
 {\cal B}_N \left ( \bm{v}_i \right ) \right |
 j = \pi(i') \right ) \nonumber \\
&=&
 \frac{1}{N-1} \sum_{i' \neq i}
 P \left ( \bm{v}_{\pi(i')}' \in
 {\cal B}_N \left ( \bm{v}_i \right )  \right ) \nonumber \\
&=&
 \frac{1}{(N-1)^2}
 \sum_{i' \neq i} E \left [ (N-1)\lambda_N (\bm{v}_i) \right ]
 \nonumber \\
&=&
\frac{1}{N-1} \int \lambda dF(p,\lambda) \nonumber \\
&=&
\frac{E[\lambda]}{N-1}
\end{eqnarray*}

%
\noindent {\it Linkage weight}: The linkage weight of the event $\bm{v}_j' \in {\cal B}_{N} \left ( \bm{v}_i\right )$ is
%
%
\begin{eqnarray*}
w &=& \log \left (\frac{P \left ( \left . \bm{v}_j' \in {\cal B}_{N} \left ( \bm{v}_i\right ) \right | M \right )}{P \left ( \left . \bm{v}_j' \in {\cal B}_{N} \left ( \bm{v}_i\right ) \right | U \right )} \right ) \nonumber \\[6pt]
&=&
\log \left (\frac{E[p]}{
E \left [ \lambda \right ]/(N-1)} \right ) \nonumber \\[6pt]
&=&
\log \left ( \frac{(N-1)E[p]}{E[\lambda]}\right ).
\end{eqnarray*}

%
\noindent {\it Conditional match probability}: The conditional match probability of the event $\bm{v}_j' \in {\cal B}_{N} \left ( \bm{v}_i\right )$ is
%
%
\begin{eqnarray*}
P \left ( M \left | \bm{v}_j' \in {\cal B}_{N} \left ( \bm{v}_i\right ) \right . \right )
 &=&
 \left ( 1 + \left ( \frac{1}{P(M)} - 1\right )
\frac{P \left ( \left . \bm{v}_j' \in {\cal B}_{N} \left ( \bm{v}_i\right ) \right | U \right )}{P \left ( \left . \bm{v}_j' \in {\cal B}_{N} \left ( \bm{v}_i\right ) \right | M \right )} 
 \right )^{-1} \nonumber \\[6pt]
 &=&
 \left ( 1 + \left ( \frac{1}{P(M)} - 1\right )
 \frac{E[\lambda]}{(N-1) E[p]} \right )^{-1} \nonumber \\[6pt]
 &=&
 \left ( 1 + \frac{E[\lambda]}{E[p]} \right )^{-1} 
\end{eqnarray*}

%
\noindent {\it Conditional Probability that the matched record is among the neighbours}: For record $i$ such that $n_i>0$, we have
{\small
%
\begin{eqnarray*}
P \left ( n_{i|M}=1 \left | n_i \right . \right ) &=& \\
\frac{E \left [ p_N \left ( \bm{v}_i \right ) {N-1 \choose n_i-1} \lambda_N(\bm{v}_i)^{n_i-1} \left ( 1 - \lambda_N(\bm{v}_i)\right )^{N-1-(n_i-1)} \right ]}{ E \left [
\lambda_N(\bm{v}_i)^{n_i-1}
\left ( 1 - \lambda_N(\bm{v}_i)\right )^{N-1-n_i}
\left ( p_N \left ( \bm{v}_i \right ) {N-1 \choose n_i-1} \left ( 1 - \lambda_N(\bm{v}_i)\right )+
\left ( 1 - p_N \left ( \bm{v}_i \right ) \right ) {N-1 \choose n_i} \lambda_N(\bm{v}_i) \right ) \right ]} & &
\end{eqnarray*}}
When $N$ is large, by the Poisson approximation\cite[Theorem 23.2]{billingsley_1995}, we have
%
%
\begin{eqnarray*}
{N-1 \choose n_i-1} \lambda_N(\bm{v}_i)^{n_i-1} \left ( 1 - \lambda_N(\bm{v}_i)\right )^{N-1-(n_i-1)} &\approx&
e^{-(N-1)\lambda_N(\bm{v}_i)}
\frac{\left ((N-1)\lambda_N(\bm{v}_i) \right )^{n_i-1}}{(n_i-1)!} \\
{N-1 \choose n_i} \lambda_N(\bm{v}_i)^{n_i} \left ( 1 - \lambda_N(\bm{v}_i)\right )^{N-1-n_i} &\approx&
e^{-(N-1)\lambda_N(\bm{v}_i)}
\frac{\left ((N-1)\lambda_N(\bm{v}_i) \right )^{n_i}}{n_i!}
\end{eqnarray*}
Hence
%
%
\begin{equation*}
P \left ( n_{i|M}=1 \left | n_i \right . \right ) \approx
\frac{E \left [ p e^{-\lambda} \frac{\lambda^{n_i-1}}{(n_i-1)!}  \right ]}{ E \left [ e^{-\lambda} \frac{\lambda^{n_i-1}}{(n_i-1)!} \left ( p + (1-p) \frac{\lambda}{n_i}  \right ) \right ]}
\end{equation*}
where $E \left [ p e^{-\lambda} \frac{\lambda^{n_i-1}}{(n_i-1)!}  \right ] = \int p e^{-\lambda} \frac{\lambda^{n_i-1}}{(n_i-1)!}  dF$, and
$E \left [ (1-p) e^{-\lambda} \frac{\lambda^{n_i}}{n_i!} \right ]=\int (1-p) e^{-\lambda} \frac{\lambda^{n_i}}{n_i!} dF$.
%

\section{Proof of Theorem 1}

\noindent Without losing any generality assume that $\alpha_{(g)}$ and $\alpha_{(g)}'$ are positive for each $g$.
Following Teicher\cite{teicher_1963}, we prove this result through Moment Generating Functions (MGFs).
%
%
\begin{eqnarray}
E \left [ z^{Y}\right ] &=&
\sum_{g=1}^G \alpha_g \left ( 1 + p_g(z-1)\right ) e^{\lambda_g (z-1)} \nonumber \\
&=&
\label{eq: mgf Y}
\sum_{g=1}^{G_0}
\alpha_{(g)} \left ( 1 + p_{(g)}(z-1)\right ) e^{\lambda_{(g)} (z-1)} \\
E \left [ z^{Y'}\right ] &=&
\sum_{g=1}^{G'} \alpha_g' \left ( 1 + p_g'(z-1) \right ) e^{\lambda_g' (z-1)} \nonumber \\
&=&
\label{eq: mgf Y'}
\sum_{g=1}^{G_0'} \alpha_{(g)}' \left ( 1 + p_{(g)}'(z-1)\right ) e^{\lambda_{(g)}' (z-1)}
\end{eqnarray}
The proof of sufficiency is straightforward.
 The condition of the theorem implies
%
%
\begin{equation}
\label{eq: two equal mgfs}
\sum_{g=1}^{G_0} \alpha_{(g)} \left ( 1 + p_{(g)}(z-1)\right ) e^{\lambda_{(g)} (z-1)} =
\sum_{g=1}^{G_0'} \alpha_{(g)}' \left ( 1 + p_{(g)}'(z-1)\right ) e^{\lambda_{(g)}' (z-1)}
\end{equation}
 and the equality of the MGFs according to Eqs.~(\ref{eq: mgf Y}) and (\ref{eq: mgf Y'}). Thus $Y$ and $Y'$ have the same distribution.

\vspace{6pt}


\noindent To prove the sufficiency, first proceed by induction to show that $\lambda_{(g)}=\lambda_{(g)}'$, $\alpha_{(g)}=\alpha_{(g)}'$ and $p_{(g)} \alpha_{(g)} = p_{(g)}' \alpha_{(g)}'$ for $g=1,\ldots,\min \left ( G_0,G_0' \right )$.
For $g=1$, start from Eq.~(\ref{eq: two equal mgfs}) and consider different cases depending on which of $p_{(1)}$ or $p_{(1)}'$ is positive.
If both are positive, we have
%
%
\begin{eqnarray*}
\lim_{z \rightarrow \infty}
\frac{\sum_{g=1}^{G_0} \alpha_{(g)}
\left ( 1 + p_{(g)}(z-1)\right )
e^{\lambda_{(g)} (z-1)}}{(z-1)e^{ \max \left (\lambda_{(1)},\lambda_{(1)}' \right ) (z-1)}}
 &=&
p_{(1)} \alpha_{(1)} I \left ( \lambda_{(1)} =
 \max \left (\lambda_{(1)},\lambda_{(1)}' \right ) \right ) \\ [6pt]
\lim_{z \rightarrow \infty}
\frac{\sum_{g=1}^{G_0'} \alpha_{(g)}'
\left ( 1 + p_{(g)}'(z-1)\right )
e^{\lambda_{(g)}' (z-1)}}{(z-1)e^{ \max \left (\lambda_{(1)},\lambda_{(1)}' \right ) (z-1)}}
 &=&
 p_{(1)}' \alpha_{(1)}' I \left ( \lambda_{(1)}' =
 \max \left (\lambda_{(1)},\lambda_{(1)}' \right ) \right )
\end{eqnarray*}
Since the MGFs are equal
%
%
$$p_{(1)} \alpha_{(1)} I \left ( \lambda_{(1)} =
 \max \left (\lambda_{(1)},\lambda_{(1)}' \right ) \right ) =
p_{(1)}' \alpha_{(1)}' I \left ( \lambda_{(1)}' =
 \max \left (\lambda_{(1)},\lambda_{(1)}' \right ) \right )$$
The above equation is incompatible with $\lambda_{(1)} \neq \lambda_{(1)}'$, because one side\footnote{the equation left hand side if $\lambda_{(1)} < \lambda_{(1)}'$ and the right hand side if $\lambda_{(1)}' < \lambda_{(1)}$} of the equation would be null while the other side\footnote{the equation right hand side if $\lambda_{(1)} < \lambda_{(1)}'$ and the left hand side if $\lambda_{(1)}' < \lambda_{(1)}$} would be positive.
Thus $\lambda_{(1)}=\lambda_{(1)}'$ and $p_{(1)} \alpha_{(1)} = p_{(1)}' \alpha_{(1)}'$ and
%
%
\begin{eqnarray*}
\sum_{g=2}^{G_0} \alpha_{(g)}
\left ( 1 + p_{(g)}(z-1)\right ) e^{\lambda_{(g)} (z-1 )} +
 \alpha_{(1)} e^{\lambda_{(1)} (z-1)} &=& \nonumber \\[6pt]
%
\sum_{g=2}^{G_0'} \alpha_{(g)}'
\left ( 1 + p_{(g)}'(z-1)\right )
e^{\lambda_{(g)}' (z-1)} +
 \alpha_{(1)}' e^{\lambda_{(1)} (z-1)} & &
\end{eqnarray*}
Consequently
%
%
\begin{eqnarray*}
\underbrace{\lim_{z \rightarrow \infty}  e^{-\lambda_{(1)} (z-1)}
\left ( \sum_{g=2}^{G_0} \alpha_{(g)}
\left ( 1 + p_{(g)}(z-1)\right ) e^{\lambda_{(g)} (z-1 )} +
 \alpha_{(1)} e^{\lambda_{(1)} (z-1)} \right ) }_{=\alpha_{(1)}} &=& \nonumber \\[6pt]
%
\underbrace{\lim_{z \rightarrow \infty}  e^{-\lambda_{(1)} (z-1)}
\left ( \sum_{g=2}^{G_0'} \alpha_{(g)}'
\left ( 1 + p_{(g)}'(z-1)\right )
e^{\lambda_{(g)}' (z-1)} +
 \alpha_{(1)}' e^{\lambda_{(1)} (z-1)} \right )}_{= \alpha_{(1)}'} & &
\end{eqnarray*}
 i.e. $\alpha_{(1)}=\alpha_{(1)}'$.


\noindent When $p_{(1)} = p_{(1)}'=0$, we obviously have $p_{(1)}\alpha_{(1)} = p_{(1)}' \alpha_{(1)}'$, which implies
%
%
\begin{eqnarray*}
\sum_{g=2}^{G_0} \alpha_{(g)}
\left ( 1 + p_{(g)}(z-1)\right ) e^{\lambda_{(g)} (z-1 )} +
 \alpha_{(1)} e^{\lambda_{(1)} (z-1)} &=& \nonumber \\[6pt]
\sum_{g=2}^{G_0'} \alpha_{(g)}'
\left ( 1 + p_{(g)}'(z-1)\right )
e^{\lambda_{(g)}' (z-1)} +
 \alpha_{(1)}' e^{\lambda_{(1)}' (z-1)} & &
\end{eqnarray*}
and
%
%
\begin{eqnarray*}
\lim_{z \rightarrow \infty}
\frac{\sum_{g=1}^{G_0} \alpha_{(g)}
\left ( 1 + p_{(g)}(z-1)\right )
e^{\lambda_{(g)} (z-1)}}{e^{ \max \left (\lambda_{(1)},\lambda_{(1)}' \right ) (z-1)}}
 &=&
 \alpha_{(1)} I \left ( \lambda_{(1)} =
 \max \left (\lambda_{(1)},\lambda_{(1)}' \right ) \right ) \\ [6pt]
\lim_{z \rightarrow \infty}
\frac{\sum_{g=1}^{G_0'} \alpha_{(g)}'
\left ( 1 + p_{(g)}'(z-1)\right )
e^{\lambda_{(g)}' (z-1)}}{e^{ \max \left (\lambda_{(1)},\lambda_{(1)}' \right ) (z-1)}}
 &=&
 \alpha_{(1)}' I \left ( \lambda_{(1)}' =
 \max \left (\lambda_{(1)},\lambda_{(1)}' \right ) \right )
\end{eqnarray*}
Thus
$$\alpha_{(1)} I \left ( \lambda_{(1)} =
 \max \left (\lambda_{(1)},\lambda_{(1)}' \right ) \right ) =
 \alpha_{(1)}' I \left ( \lambda_{(1)}' =
 \max \left (\lambda_{(1)},\lambda_{(1)}' \right ) \right )$$
 which is incompatible with $\lambda_{(1)} \neq \lambda_{(1)}'$. Thus
$\lambda_{(1)} = \lambda_{(1)}'$ and $\alpha_{(1)} =\alpha_{(1)}'$.


\noindent To complete the proof of the property for $g=1$, we next show that we can neither have $p_{(1)}=0$ while $p_{(1)}'>0$ nor $p_{(1)}>0$ while $p_{(1)}'=0$. Indeed, suppose that  $p_{(1)}=0$ and $p_{(1)}'>0$.
%
%
\begin{eqnarray*}
\lim_{z \rightarrow \infty}
\frac{\sum_{g=1}^{G_0} \alpha_{(g)}
\left ( 1 + p_{(g)}(z-1)\right )
e^{\lambda_{(g)} (z-1)}}{(z-1) e^{ \max \left (\lambda_{(1)},\lambda_{(1)}' \right ) (z-1)}}
 &=& 0 \\ [6pt]
\lim_{z \rightarrow \infty}
\frac{\sum_{g=1}^{G_0'} \alpha_{(g)}'
\left ( 1 + p_{(g)}'(z-1)\right )
e^{\lambda_{(g)}' (z-1)}}{(z-1) e^{ \max \left (\lambda_{(1)},\lambda_{(1)}' \right ) (z-1)}}
 &=&
 \alpha_{(1)}' I \left ( \lambda_{(1)}' =
 \max \left (\lambda_{(1)},\lambda_{(1)}' \right ) \right )
\end{eqnarray*}
 which implies $\lambda_{(1)}' < \lambda_{(1)}$ and
\begin{eqnarray*}
\lim_{z \rightarrow \infty}
\frac{\sum_{g=1}^{G_0} \alpha_{(g)}
\left ( 1 + p_{(g)}(z-1)\right )
e^{\lambda_{(g)} (z-1)}}{e^{ \lambda_{(1)} (z-1)}}
 &=& \alpha_{(1)} \\ [6pt]
\lim_{z \rightarrow \infty}
\frac{\sum_{g=1}^{G_0'} \alpha_{(g)}'
\left ( 1 + p_{(g)}'(z-1)\right )
e^{\lambda_{(g)}' (z-1)}}{e^{ \lambda_{(1)} (z-1)}}
 &=&
 0
\end{eqnarray*}
 The equality of the MGFs implies the equality of the two limits and the contradiction $\alpha_{(1)}=0$.
 If $p_{(1)}>0$ while $p_{(1)}'=0$, we can prove in a similar manner that this implies the contradiction $\alpha_{(1)}'=0$. Thus the induction property is true for $g=1$. 


\noindent Now suppose that the property holds from $1$ up to $t < \min (G_0,G_0')$. Since the MGFs are equal and the property holds up to $t$, we have
$$\sum_{g=t+1}^{G_0} \alpha_{(g)}
  \left ( 1 + p_{(g)}(z-1)\right ) e^{\lambda_{(g)}(z-1)} =
 \sum_{g=t+1}^{G_0'} \alpha_{(g)}'
 \left ( 1 + p_{(g)}'(z-1)\right ) e^{\lambda_{(g)}'(z-1)},
 \ \forall z$$
Next consider different cases depending on which of $p_{(t+1)}$ or $p_{(t+1)}'$ is positive.
 If both are positive, we have
\begin{eqnarray*}
\lim_{z \rightarrow \infty}
\frac{\sum_{g=t+1}^{G_0} \alpha_{(g)}
 \left ( 1 + p_{(g)}(z-1)\right ) e^{\lambda_{(g)}(z-1)}}{(z-1)e^{ \max \left (\lambda_{(t+1)},\lambda_{(t+1)}' \right ) (z-1)}} &=&
 p_{(t+1)} \alpha_{(t+1)} I \left ( \lambda_{(t+1)} = \max \left (\lambda_{(t+1)},\lambda_{(t+1)}' \right ) \right ) \\ [6pt]
\lim_{z \rightarrow \infty}
\frac{\sum_{g=t+1}^{G_0} \alpha_{(g)}'
 \left ( 1 + p_{(g)}'(z-1)\right ) e^{\lambda_{(g)}'(z-1)}}{(z-1)e^{ \max \left (\lambda_{(t+1)},
\lambda_{(t+1)}' \right ) (z-1)}} &=&
 p_{(t+1)}' \alpha_{(t+1)}' I \left ( \lambda_{(t+1)}' = \max \left (\lambda_{(t+1)},\lambda_{(t+1)}' \right ) \right ) 
\end{eqnarray*}
 Thus
$$ p_{(t+1)} \alpha_{(t+1)} I \left ( \lambda_{(t+1)} = \max \left (\lambda_{(t+1)},\lambda_{(t+1)}' \right ) \right ) =
 p_{(t+1)}' \alpha_{(t+1)}' I \left ( \lambda_{(t+1)}' = \max \left (\lambda_{(t+1)},\lambda_{(t+1)}' \right ) \right )$$
 which implies $\lambda_{(t+1)}=\lambda_{(t+1)}'$,
 $p_{(t+1)} \alpha_{(t+1)}=p_{(t+1)}' \alpha_{(t+1)}'$, and
\begin{eqnarray*}
\underbrace{\lim_{z \rightarrow \infty} 
e^{-\lambda_{(t+1)} (z-1)} \left ( \sum_{g=t+2}^{G_0} \alpha_{(g)}
\left ( 1 + p_{(g)}(z-1)\right ) e^{\lambda_{(g)} (z-1 )} +
 \alpha_{(t+1)} e^{\lambda_{(t+1)} (z-1)} \right )}_{=\alpha_{(t+1)}} &=& \nonumber \\[6pt]
\underbrace{\lim_{z \rightarrow \infty}  e^{-\lambda_{(t+1)} (z-1)} \left ( \sum_{g=t+2}^{G_0'} \alpha_{(g)}'
\left ( 1 + p_{(g)}'(z-1)\right ) e^{\lambda_{(g)}' (z-1)} +
 \alpha_{(t+1)}' e^{\lambda_{(t+1)} (z-1)} \right )}_{= \alpha_{(t+1)}'} & &
\end{eqnarray*}
i.e. $\alpha_{(t+1)}=\alpha_{(t+1)}'$, and the property holds.

\noindent When $p_{(t+1)} = p_{(t+1)}'=0$, we obviously have $p_{(t+1)}\alpha_{(t+1)} = p_{(t+1)}' \alpha_{(t+1)}'$, which implies
%
%
\begin{eqnarray*}
\sum_{g=t+1}^{G_0} \alpha_{(g)}
\left ( 1 + p_{(g)}(z-1)\right ) e^{\lambda_{(g)} (z-1 )} +
 \alpha_{(t+1)} e^{\lambda_{(t+1)} (z-1)} &=& \nonumber \\[6pt]
\sum_{g=t+1}^{G_0'} \alpha_{(g)}'
\left ( 1 + p_{(g)}'(z-1)\right )
e^{\lambda_{(g)}' (z-1)} +
 \alpha_{(t+1)}' e^{\lambda_{(t+1)}' (z-1)} & &
\end{eqnarray*}
and
%
%
\begin{eqnarray*}
\lim_{z \rightarrow \infty}
\frac{\sum_{g=t+1}^{G_0} \alpha_{(g)}
\left ( 1 + p_{(g)}(z-1)\right )
e^{\lambda_{(g)} (z-1)}}{e^{ \max \left (\lambda_{(t+1)},\lambda_{(t+11)}' \right ) (z-1)}}
 &=&
 \alpha_{(t+1)} I \left ( \lambda_{(t+1)} =
 \max \left (\lambda_{(t+1)},\lambda_{(t+1)}' \right ) \right ) \\ [6pt]
\lim_{z \rightarrow \infty}
\frac{\sum_{g=t+1}^{G_0'} \alpha_{(g)}'
\left ( 1 + p_{(g)}'(z-1)\right )
e^{\lambda_{(g)}' (z-1)}}{e^{ \max \left (\lambda_{(t+1)},\lambda_{(t+1)}' \right ) (z-1)}}
 &=&
 \alpha_{(t+1)}' I \left ( \lambda_{(t+1)}' =
 \max \left (\lambda_{(t+1)},\lambda_{(t+1)}' \right ) \right )
\end{eqnarray*}
Thus
$$\alpha_{(t+1)} I \left ( \lambda_{(t+1)} =
 \max \left (\lambda_{(t+1)},\lambda_{(t+1)}' \right ) \right ) =
 \alpha_{(t+1)}' I \left ( \lambda_{(t+1)}' =
 \max \left (\lambda_{(t+1)},\lambda_{(t+1)}' \right ) \right )$$
 which is incompatible with $\lambda_{(t+1)} \neq \lambda_{(t+1)}'$. Thus
$\lambda_{(t+1)} = \lambda_{(t+1)}'$,
 $\alpha_{(t+1)} =\alpha_{(t+1)}'$ and the property holds


\noindent To conclude, we next show that we cannot have $p_{(t+1)}=0$ and $p_{(t+1)}'>0$ or $p_{(t+1)}>0$ and $p_{(t+1)}'=0$.
We only consider the case $p_{(t+1)}=0$ and $p_{(t+1)}'>0$ since the proof is similar for the other case.
If $p_{(t+1)}=0$ and $p_{(t+1)}'>0$, we have
%
%
\begin{eqnarray*}
\lim_{z \rightarrow \infty}
\frac{\sum_{g=t+1}^{G_0} \alpha_{(g)}
\left ( 1 + p_{(g)}(z-1)\right )
e^{\lambda_{(g)} (z-1)}}{(z-1)e^{ \max \left (\lambda_{(t+1)},\lambda_{(t+1)}' \right ) (z-1)}}
 &=& 0 \\ [6pt]
\lim_{z \rightarrow \infty}
\frac{\sum_{g=t+1}^{G_0'} \alpha_{(g)}'
\left ( 1 + p_{(g)}'(z-1)\right )
e^{\lambda_{(g)}' (z-1)}}{(z-1)e^{ \max \left (\lambda_{(t+1)},\lambda_{(t+1)}' \right ) (z-1)}}
 &=&
 \alpha_{(t+1)}' I \left ( \lambda_{(t+1)}' =
 \max \left (\lambda_{(t+1)},\lambda_{(t+1)}' \right ) \right )
\end{eqnarray*}
 which implies that $\lambda_{(t+1)}' < \lambda_{(t+1)}$ and the following limits
%
%
\begin{eqnarray*}
\lim_{z \rightarrow \infty}
\frac{\sum_{g=t+1}^{G_0} \alpha_{(g)}
\left ( 1 + \eta p_{(g)}(z-1)\right )
e^{\lambda_{(g)} (z-1)}}{e^{ \lambda_{(t+1)} (z-1)}}
 &=&  \alpha_{(t+1)} \\ [6pt]
\lim_{z \rightarrow \infty}
\frac{\sum_{g=1}^{G_0'} \alpha_{(g)}'
\left ( 1 + \eta' p_{(g)}'(z-1)\right )
e^{\lambda_{(g)}' (z-1)}}{e^{ \lambda_{(t+1)} (z-1)}}
 &=& 0
\end{eqnarray*}
 which implies $\alpha_{(t+1)}=0$; a contradiction.
 This concludes the proof of the induction property.

\noindent Next proceed by contradiction to show that $G_0=G_0'$.
Indeed $G_0<G_0'$ implies
$$
\sum_{g=t+1}^{G_0'} \alpha_{(g)}' \left ( 1 + p_{(g)}'(z-1)\right ) e^{\lambda_{(g)}'(z-1)}=0, \ \forall z
$$
including for $z>1$, which is impossible.
For a similar reason, we cannot have $G_0>G_0'$. Hence $G_0=G_0'$.
 Since $\alpha_{(g)}$ is positive, $\alpha_{(g)}=\alpha_{(g)}'$  and $\alpha_{(g)}p_{(g)} = \alpha_{(g)}' p_{(g)}'$, we have $p_{(g)} = p_{(g)}'$.
 This completes the proof.

\hfill Q.E.D.


\section{Proof of Theorem 2}

\noindent The proof is a consequence of
 Lemmas~\ref{lemma: first bound characteristic function}
 and~\ref{lemma: second bound characteristic function}.
 Lemma~\ref{lemma: first bound characteristic function} is supported by
 Lemmas~\ref{lemma: max_p_1n_p_2n} and~\ref{lemma: first bound ratio},
 while Lemma~\ref{lemma: second bound characteristic function} is supported by
 Lemmas~\ref{lemma: bound on the multinomial to poisson pmf ratio}
 and~\ref{lemma: exponential bound poisson multinomial cdf}.
 For the essence of the proof, one may first read
 the proofs of Lemma~\ref{lemma: first bound characteristic function} 
 and Lemma~\ref{lemma: second bound characteristic function},
 in sequence, before those of the supporting lemmas.

%
\noindent Let ${\cal E}_{1N}$ denote the event that $\bm{v}_{\pi(i')}'$ does not belong to ${\cal B}_N \left ( \bm{v}_i\right )$, for all $1 \leq i\neq i' \leq m_N$, i.e.
%
%
\begin{equation}
 \label{eq event E_1N}
 {\cal E}_{1N} =
 \bigcap_{i=1}^{m_N} \left \{ \bm{v}_{\pi(i)}' \in
 \bigcap_{t \in \{1,\ldots,m_N\}-\{i\}} {\cal B}_N \left ( \bm{v}_t \right )^c
 \right \}.
\end{equation}
\noindent Also let ${\cal E}_{2N}$ denote the event that the neighbourhoods
 ${\cal B}_N \left ( \bm{v}_1 \right ),\ldots,{\cal B}_N\left ( \bm{v}_{m_N}\right )$ are
 disjoint, i.e.
%
%
\begin{equation}
 \label{eq event E_2N}
 {\cal E}_{2N} =
 \bigcap_{1 \leq i < i' \leq m_N}
 \left \{ {\cal B}_N \left ( \bm{v}_i \right ) \cap
 {\cal B}_N \left ( \bm{v}_{i'} \right ) = \emptyset \right \}.
\end{equation}

\noindent The next lemma gives a lowerbound on the probabilities of these two events.
%
%
\begin{lemma} \label{lemma: max_p_1n_p_2n}

\noindent When $N$ is large enough,
\begin{equation}
\max \left ( P \left ( {\cal E}_{1N}^c \right ),
             P \left ( {\cal E}_{2N}^c \right )  \right ) \leq
\frac{2\Lambda m_N^2}{N}.
\end{equation}

\end{lemma}

\noindent \underline{Proof}: For ${\cal E}_{1N}^c$, we apply Bonferroni's inequality as follows:
%
%
\begin{eqnarray*}
 P \left ( {\cal E}_{1N}^c \right ) &=&
 P \left ( \bigcup_{i=1}^{m_N} \left \{ \bm{v}_{\pi(i)}' \in
 \bigcup_{t \in \{1,\ldots,m_N\}-\{i\}} {\cal B}_N \left ( \bm{v}_t \right )
 \right \} \right ) \\[3pt]
 &\leq&
  \sum_{i=1}^{m_N} \sum_{t \in \{1,\ldots,m_N\}-\{i\}}
  P \left ( \bm{v}_{\pi(i)}' \in {\cal B}_N \left ( \bm{v}_t \right ) \right ) \\[3pt]
 &=&
 \sum_{i=1}^{m_N} \sum_{t \in \{1,\ldots,m_N\}-\{i\}} 
 \frac{1}{N-1} E \left [ (N-1) P \left ( \left . \bm{v}_{\pi(i)}' \in
 {\cal B}_N \left ( \bm{v}_t \right ) \right | \bm{v}_t \right ) \right ] \\[3pt]
 &\leq& \frac{\Lambda m_N \left ( m_N-1\right )}{N-1} \\[3pt]
 &\leq&
 \frac{2\Lambda m_N^2}{N},
\end{eqnarray*}
 when $N$ is large enough.

\noindent For ${\cal E}_{2N}^c$, we also apply Bonferroni's inequality as follows:
%
%
\begin{eqnarray*}
 P \left ( {\cal E}_{1N}^c \right ) &=&
 P \left ( \bigcup_{1 \leq i < i' \leq m_N}
 \left \{ {\cal B}_N \left ( \bm{v}_i \right ) \cap
 {\cal B}_N \left ( \bm{v}_{i'} \right ) \neq \emptyset \right \} \right ) \\[3pt]
 &\leq&
 \sum_{1 \leq i < i' \leq m_N} P \left ( \left \{ {\cal B}_N \left ( \bm{v}_i \right ) \cap
 {\cal B}_N \left ( \bm{v}_{i'} \right ) \neq \emptyset \right \} \right ) \\[3pt]
 &\leq&
 \frac{\Lambda {m_N \choose 2} }{N} \\[3pt]
 &\leq&
 \frac{2\Lambda m_N^2}{N},
\end{eqnarray*}
 when $m_N>2$.

\hfill Q.E.D.

\noindent For $i=1,\ldots,m_N$ and $j=1,\ldots,N$,
 let $Y_{ij} = I \left ( \bm{v}_j' \in {\cal B}_N \left ( \bm{v}_i \right ) \right )$,
 $\bm{Y}_i = \left [ Y_{ij} \right ]_{1 \leq j \leq N}$ and
 $\bm{Y} = \left [ \bm{Y}_i \right ]_{1 \leq i \leq m_N}$.
%
%
\begin{lemma} \label{lemma: first bound ratio}

\noindent Let $\bm{v}$ and $\bm{y}$ be such that
 ${\cal B}_N \left ( \bm{v}_1\right ),\ldots,{\cal B}_N \left ( \bm{v}_m\right )$ are disjoint
 and $y_{it}=0$ if $i=1,\ldots,m_N$ and $t \neq i$. Then
\begin{equation}
\label{eq bound on ratio where i lt m_N}
\left | \frac{P \left ( \bm{Y}=\bm{y} \left | \bm{V}=\bm{v}, {\cal E}_{1N} \right . \right )}{
P \left ( \bm{Y}=\bm{y} \left | \bm{V}=\bm{v} \right . \right )} - 1 \right | \leq
 \left ( 1 + \frac{\Lambda}{(N-1)-\Lambda} \right )
 \frac{\Lambda m_N \left ( m_N-1 \right )}{(N-1) - \Lambda \left ( m_N-1 \right )}
\end{equation}

\end{lemma}


\noindent \underline{Proof}:
%
%
{\small
\begin{eqnarray*}
\frac{P \left ( \bm{Y}=\bm{y} \left | \bm{V}=\bm{v}, {\cal E}_{1N} \right . \right )}{
P \left ( \bm{Y}=\bm{y} \left | \bm{V}=\bm{v} \right . \right )} &=&
\nonumber \\[6pt]
\frac{P \left ( \bm{Y}=\bm{y}, {\cal E}_{1N} \left | \bm{V}=\bm{v} \right . \right )}{
P \left ( {\cal E}_{1N} \left | \bm{V}=\bm{v} \right . \right )
P \left ( \bm{Y}=\bm{y} \left | \bm{V}=\bm{v} \right . \right )}
&=& \nonumber \\[6pt]
\frac{P \left ( \left [ \bm{Y}_i \right ]_{1 \leq i \leq m_N}=
\left [ \bm{y}_i \right ]_{1 \leq i \leq m_N},
{\cal E}_{1N} \left | \bm{V}=\bm{v} \right . \right )
P \left ( \left [ \bm{Y}_i \right ]_{m_N+1 \leq i \leq N}=
\left [ \bm{y}_i \right ]_{m_N+1 \leq i \leq N} \left | \bm{V}=\bm{v} \right . \right )}{
P \left ( {\cal E}_{1N} \left | \bm{V}=\bm{v} \right . \right )
P \left ( \bm{Y}=\bm{y} \left | \bm{V}=\bm{v} \right . \right )}
&=& \nonumber \\[6pt]
\frac{P \left ( \left [ \bm{Y}_i \right ]_{1 \leq i \leq m_N}=
\left [ \bm{y}_i \right ]_{1 \leq i \leq m_N} \left |
\bm{V}=\bm{v}, {\cal E}_{1N} \right . \right )}{
P \left ( \left [ \bm{Y}_i \right ]_{1 \leq i \leq m_N}=
\left [ \bm{y}_i \right ]_{1 \leq i \leq m_N} \left | \bm{V}=\bm{v} \right . \right )}
&=& \nonumber \\[6pt]
 \prod_{i=1}^{m_N} \left ( P \left ( \left . \bm{V}_{\pi(i)}' \in
 \bigcap_{t \in \{1,\ldots,m_N\}-\{i\}} {\cal B}_N \left ( \bm{v}_t \right )^c \right |
 \bm{V}=\bm{v} \right )^{-y_{ii}} \times \right .
& & \nonumber \\[6pt]
 \left . \left ( \frac{ \displaystyle 1 - \frac{p_N \left ( \bm{v}_i \right )}{
 P \left ( \left . \bm{V}_{\pi(i)}' \in \bigcap_{t \in \{1,\ldots,m_N\}-\{i\}} {\cal B}_N \left ( \bm{v}_t \right )^c \right |
 \bm{V}=\bm{v} \right )}}{1-p_N \left ( \bm{v}_i \right )} \right )^{1-y_{ii}} \right )
 & &
\end{eqnarray*}}
 Hence
%
%
{\small
\begin{eqnarray*}
\left | \frac{P \left ( \bm{Y}=\bm{y} \left | \bm{V}=\bm{v}, {\cal E}_{1N} \right . \right )}{
P \left ( \bm{Y}=\bm{y} \left | \bm{V}=\bm{v} \right . \right )} - 1 \right |
&\leq& \nonumber \\[6pt]
 \sum_{i=1}^{m_N} y_{ii}
 \left | 1 - \frac{1}{P \left ( \left . \bm{V}_{\pi(i)}' \in
 \bigcap_{t \in \{1,\ldots,m_N\}-\{i\}} {\cal B}_N \left ( \bm{v}_t \right )^c \right |
 \bm{V}=\bm{v} \right )} \right | +
 & & \nonumber \\[6pt]
\sum_{i=1}^{m_N} (1-y_{ii})
\left | 1 - \frac{ \displaystyle
1 - \frac{p_N \left ( \bm{v}_i \right )}{
 P \left ( \left . \bm{V}_{\pi(i)}' \in
 \bigcap_{t \in \{1,\ldots,m_N\}-\{i\}} {\cal B}_N \left ( \bm{v}_t \right )^c \right |
 \bm{V}=\bm{v} \right )}}{1-p_N \left ( \bm{v}_i \right )} \right |
&=& \nonumber \\[6pt]
 \sum_{i=1}^{m_N}
 \left ( y_{ii} + \left ( 1 - y_{ii} \right )
 \frac{p_N \left ( \bm{v}_i \right )}{1-p_N \left ( \bm{v}_i \right )} \right )
 \left | 1 - \frac{1}{P \left ( \left . \bm{V}_{\pi(i)}' \in
 \bigcap_{t \in \{1,\ldots,m_N\}-\{i\}} {\cal B}_N \left ( \bm{v}_t \right )^c \right |
 \bm{V}=\bm{v} \right )} \right |
&\leq& \nonumber \\[6pt]
 \sum_{i=1}^{m_N}
 \left ( 1 + \frac{\Lambda}{(N-1)-\Lambda} \right )
 \left | 1 - \frac{1}{1 - P \left ( \left . \bm{V}_{\pi(i)}' \in
 \bigcup_{t \in \{1,\ldots,m_N\}-\{i\}} {\cal B}_N \left ( \bm{v}_t \right ) \right |
 \bm{V}=\bm{v} \right )} \right |
&\leq& \nonumber \\[6pt]
 \left ( 1 + \frac{\Lambda}{(N-1)-\Lambda} \right )
 \sum_{i=1}^{m_N}
 \left | 1 - \frac{1}{1 - \sum_{t \in \{1,\ldots,m_N\}-\{i\}}
 P \left ( \left . \bm{V}_{\pi(i)}' \in {\cal B}_N \left ( \bm{v}_t \right ) \right |
 \bm{V}=\bm{v} \right )} \right |
&=& \nonumber \\[6pt]
 \left ( 1 + \frac{\Lambda}{(N-1)-\Lambda} \right )
 \sum_{i=1}^{m_N}
 \left | \frac{\sum_{t \in \{1,\ldots,m_N\}-\{i\}}
 P \left ( \left . \bm{V}_{\pi(i)}' \in {\cal B}_N \left ( \bm{v}_t \right ) \right |
 \bm{V}_i=\bm{v}_i, \bm{V}_t=\bm{v}_t\right )}{1 - \sum_{t \in \{1,\ldots,m_N\}-\{i\}}
 P \left ( \left . \bm{V}_{\pi(i)}' \in {\cal B}_N \left ( \bm{v}_t \right ) \right |
 \bm{V}_i=\bm{v}_i, \bm{V}_t=\bm{v}_t\right )} \right |
&\leq& \nonumber \\[6pt]
 \left ( 1 + \frac{\Lambda}{(N-1)-\Lambda} \right )
 \frac{\Lambda m_N \left ( m_N-1 \right )}{(N-1) - \Lambda \left ( m_N-1 \right )},
& &
\end{eqnarray*}}
 where the last equation follows from  (\ref{eq bound on other capture probability}) because
 ${\cal B}_N \left (\bm{v}_1\right ),\ldots,{\cal B}_N \left ( \bm{v}_{m_N}\right )$ are disjoint.

\hfill Q.E.D.

\vspace{6pt}

%
\begin{lemma} \label{lemma: first bound characteristic function}
 Let $\tilde{\bm{V}}=\left [ \tilde{\bm{V}}_t \right ]_{1 \leq t \leq m_N}$,
 $\tilde{\bm{Y}}=\left [ \tilde{\bm{Y}}_i \right ]_{1 \leq i \leq N}$ and
 $\tilde{\bm{n}} = \sum_{i=1}^N \tilde{\bm{Y}}_i$ be such that
 $\tilde{\bm{V}} \stackrel{d}{=} \bm{V}$,
 $\tilde{\bm{Y}}_1,\ldots, \tilde{\bm{Y}}_N$ are mutually independent given $\tilde{\bm{V}}$,
 $$\left . \left ( \tilde{\bm{Y}}_{m_N+1},\ldots,
 \tilde{\bm{Y}}_N \right ) \right |
 \left \{ \tilde{\bm{V}} = \tilde{\bm{v}} \right \} \stackrel{d}{=}
   \left . \left ( \bm{Y}_{m_N+1},\ldots, \bm{Y}_N \right ) \right |
 \left \{ \bm{V} = \tilde{\bm{v}} \right \},$$
 for $i=1,\ldots,m_N$, and
%
%
\begin{eqnarray}
 \left . \tilde{Y}_{ii} \right |
 \left \{ \tilde{\bm{V}} = \tilde{\bm{v}} \right \} &\stackrel{d}{=}&
 \left . Y_{ii} \right | \left \{ \bm{V} = \tilde{\bm{v}} \right \}, \\[3pt]
 \tilde{Y}_{it} &=& 0, \ t \neq i.
\end{eqnarray}
 Then for any $\bm{T}_N: I\!\!N^{m_N} \rightarrow I\!\!R^d$ and
 $\bm{\omega} \in I\!\!R^d$
%
%
%
\begin{equation}
 \left | E \left [ e^{\jmath \bm{\omega}^{\top} \bm{T}_N \left ( \bm{n} \right ) } \right ] -
 E \left [  e^{\jmath \bm{\omega}^{\top} \bm{T}_N \left ( \tilde{\bm{n}} \right ) } \right ]
 \right | =
 O \left (m_N^2/N \right ),
\end{equation}
 where $N$ is large enough and $\jmath^2 = -1$.
\end{lemma}


\noindent \underline{Proof}:
We have
%
%
\begin{eqnarray}
 E \left [ e^{\jmath \bm{\omega}^{\top} \bm{T}_N \left ( \bm{n} \right ) } \right ] &=&
 E \left [ \left .  I \left ( {\cal E}_{2N} \right )
 e^{\jmath \bm{\omega}^{\top} \bm{T}_N \left ( \bm{n} \right ) } \right | {\cal E}_{1N} \right ] +
 E \left [ I \left ( {\cal E}_{2N}^c \right ) e^{\jmath \bm{\omega}^{\top} \bm{T}_N \left ( \bm{n} \right ) } \right ] + \nonumber \\[6pt]
 & &
 P \left ( {\cal E}_{1N}^c\right ) \left (
 E \left [ \left .  I \left ( {\cal E}_{2N} \right )
 e^{\jmath \bm{\omega}^{\top} \bm{T}_N \left ( \bm{n} \right ) } \right | {\cal E}_{1N}^c \right ] -
 E \left [ \left .  I \left ( {\cal E}_{2N} \right )
 e^{\jmath \bm{\omega}^{\top} \bm{T}_N \left ( \bm{n} \right ) } \right | {\cal E}_{1N} \right ] \right ) \nonumber \\
\end{eqnarray}
Using a change of measure, we have
%
%
\begin{equation}
 E \left [ \left .  I \left ( {\cal E}_{2N} \right )
 e^{\jmath \bm{\omega}^{\top} \bm{T}_N \left ( \bm{n} \right ) } \right | {\cal E}_{1N} \right ] =
 E \left [  I \left ( \tilde{\cal E}_{2N}  \right )
 \tilde{r} \left ( \tilde{\bm{V}}, \tilde{\bm{Y}} \right )
 e^{\jmath \bm{\omega}^{\top} \bm{T}_N \left ( \tilde{\bm{n}} \right ) } \right ],
\end{equation}
where
%
%
\begin{eqnarray}
 \tilde{r} \left ( \tilde{\bm{v}}, \tilde{\bm{y}} \right ) &=&
 \frac{P \left ( \bm{Y}=\tilde{\bm{y}} \left |
 \bm{V}=\tilde{\bm{v}}, {\cal E}_{1N} \right . \right )}{
 P \left ( \bm{Y}=\tilde{\bm{y}} \left | \bm{V}=\tilde{\bm{v}} \right . \right )}, \\
%
%
 \tilde{{\cal E}}_{2N} &=&
 \bigcap_{1 \leq i < i' \leq m_N}
 \left \{ {\cal B}_N \left ( \tilde{\bm{V}}_i \right ) \cap
 {\cal B}_N \left ( \tilde{\bm{V}}_{i'} \right ) = \emptyset \right \}.
\end{eqnarray}
Hence
%
%
\begin{eqnarray}
 E \left [ e^{\jmath \bm{\omega}^{\top} \bm{T}_N \left ( \bm{n} \right ) } \right ] &=&
 E \left [ I \left ( \tilde{\cal E}_{2N} \right )
 e^{\jmath \bm{\omega}^{\top} \bm{T}_N \left ( \tilde{\bm{n}} \right ) } \right ] +
 E \left [  I \left ( \tilde{\cal E}_{2N}  \right )
 \left ( \tilde{r} \left ( \tilde{\bm{V}}, \tilde{\bm{Y}} \right ) - 1 \right )
 e^{\jmath \bm{\omega}^{\top} \bm{T}_N \left ( \tilde{\bm{n}} \right ) } \right ] +
 \nonumber \\[6pt]
 & &
 E \left [ I \left ( {\cal E}_{2N}^c \right )
 e^{\jmath \bm{\omega}^{\top} \bm{T}_N \left ( \bm{n} \right ) } \right ] +
 \nonumber \\[6pt]
 & &
 P \left ( {\cal E}_{1N}^c\right ) \left (
 E \left [ \left .  I \left ( {\cal E}_{2N} \right )
 e^{\jmath \bm{\omega}^{\top} \bm{T}_N \left ( \bm{n} \right ) } \right | {\cal E}_{1N}^c \right ] -
 E \left [ \left .  I \left ( {\cal E}_{2N} \right )
 e^{\jmath \bm{\omega}^{\top} \bm{T}_N \left ( \bm{n} \right ) } \right |
 {\cal E}_{1N} \right ] \right ) \nonumber \\[6pt]
%
%
 &=&
 E \left [ e^{\jmath \bm{\omega}^{\top} \bm{T}_N \left ( \tilde{\bm{n}} \right ) } \right ] -
 E \left [ I \left ( \tilde{\cal E}_{2N}^{c} \right )
 e^{\jmath \bm{\omega}^{\top} \bm{T}_N \left ( \tilde{\bm{n}} \right ) } \right ] +
 \nonumber \\[6pt]
 & &
 E \left [  I \left ( \tilde{\cal E}_{2N}  \right )
 \left ( \tilde{r} \left ( \tilde{\bm{V}}, \tilde{\bm{Y}} \right ) - 1 \right )
 e^{\jmath \bm{\omega}^{\top} \bm{T}_N \left ( \tilde{\bm{n}} \right ) } \right ] +
 \nonumber \\[6pt]
 & &
 E \left [ I \left ( {\cal E}_{2N}^c \right )
 e^{\jmath \bm{\omega}^{\top} \bm{T}_N \left ( \bm{n} \right ) } \right ] +
 \nonumber \\[6pt]
 & &
 P \left ( {\cal E}_{1N}^c\right ) \left (
 E \left [ \left .  I \left ( {\cal E}_{2N} \right )
 e^{\jmath \bm{\omega}^{\top} \bm{T}_N \left ( \bm{n} \right ) } \right |
 {\cal E}_{1N}^c \right ] - \right . \nonumber \\[6pt]
 & &
 \left . E \left [ \left .  I \left ( {\cal E}_{2N} \right )
 e^{\jmath \bm{\omega}^{\top} \bm{T}_N \left ( \bm{n} \right ) } \right |
 {\cal E}_{1N} \right ] \right ).
\end{eqnarray}
 According to Lemma~\ref{lemma: first bound ratio} (Eq.~(\ref{eq bound on ratio where i lt m_N})),
 we have
%
%
\begin{equation}
 I \left ( \tilde{\cal E}_{2N}  \right )\left |
 \tilde{r} \left ( \tilde{\bm{V}}, \tilde{\bm{Y}} \right ) - 1 \right | \leq 
 \left ( 1 + \frac{\Lambda}{(N-1)-\Lambda} \right )
 \frac{\Lambda m_N \left ( m_N-1 \right )}{(N-1) - \Lambda \left ( m_N-1 \right )}.
\end{equation}
 When $N$ is large enough
%
%
\begin{eqnarray*}
 \left | E \left [ e^{\jmath \bm{\omega}^{\top} \bm{T}_N \left ( \bm{n} \right ) } \right ] -
 E \left [ e^{\jmath \bm{\omega}^{\top} \bm{T}_N \left ( \tilde{\bm{n}} \right ) } \right ]
 \right |
 &\leq&
 2P \left ( {\cal E}_{1N}^c\right ) + P\left ( \tilde{\cal E}_{2N}^c \right ) +
 P \left ( {\cal E}_{2N}^c\right ) + \frac{2\Lambda m_N^2}{N} \\
 &=&
 2P \left ( {\cal E}_{1N}^c\right ) + 2P \left ( {\cal E}_{2N}^c\right ) +
 \frac{2\Lambda m_N^2}{N},
\end{eqnarray*}
 because $P\left ( \tilde{\cal E}_{2N}^{c} \right ) = P \left ( {\cal E}_{2N}^c\right )$.
 To conclude, note that from Lemma~\ref{lemma: max_p_1n_p_2n}, we have
$$2P \left ( {\cal E}_{1N}^c\right ) + 2P \left ( {\cal E}_{2N}^c\right ) =
 O \left ( m_N^2/N\right ).$$

\hfill Q.E.D.

%
\begin{lemma} \label{lemma: bound on the multinomial to poisson pmf ratio}
 \noindent Let
 $\bm{\lambda}=
  \left [ \lambda_t \right ]_{1 \leq t \leq m_N}=
  \left [ \lambda_N \left ( \bm{v}_t\right ) \right ]_{1 \leq t \leq m_N}$,
 $\bm{Z} = \left [ Z_t \right ]_{1 \leq t \leq m_N}$ such that
 $$\left ( N-m_N- \sum_{t=1}^{m_N} Z_t, \bm{Z} \right )
   \sim \mbox{multinomial}\left ( N-m_N,
   \left ( 1-\sum_{t=1}^{m_N} \lambda_t, \bm{\lambda} \right )\right ),$$
 $b_N = O \left ( m_N \right )$ and
 $\bm{z} = \left [ z_t \right ]_{1 \leq t \leq m_N}$ be such that
 $\sum_{t=1}^{m_N} z_t \leq b_N$.
 Then
%
%
\begin{equation}
\left ( \prod_{t=1}^{m_N} \frac{e^{-(N-1)\lambda_t}
\left ((N-1)\lambda_t \right )^{z_t}}{z_t!}
\right )^{-1}P \left ( \bm{Z}=\bm{z}\right )=
\exp \left ( O \left ( m_N^2/{N}\right ) \right )
\end{equation}
\end{lemma}
%
%
\noindent \underline{Proof}:
 Let $b_N = o \left ( N/m_N \right )$.
 Using the connection between the multinomial and
 Poisson sampling models \cite[chap. 2.1.4]{agresti_2002}, we have
%
%
\begin{eqnarray*}
 P \left ( \bm{Z}=\bm{z} \right ) &=& \\
 P \left ( \left ( N-m_N - \sum_{t=1}^{m_N} Z_t, \bm{Z} \right ) =
 \left ( N-m_N - \sum_{t=1}^{m_N} z_t, \bm{z} \right )\right ) &=& \\
\frac{ e^{-(N-m_N)\left ( 1-\sum_{t=1}^{m_N} \lambda_t \right )} \frac{\left ( (N-m_N)\left ( 1 - \sum_{t=1}^{m_N} \lambda_t \right )\right )^{N-m_N - \sum_{t=1}^{m_N} z_t }}{\left ( N-m_N - \sum_{t=1}^{m_N} z_t \right )!}
\prod_{t=1}^{m_N} \left ( e^{-(N-m_N)\lambda_t} \frac{\left ( (N-m_N)\lambda_t\right )^{z_t}}{z_t!} \right ) }{ e^{-(N-m_N)} \frac{ (N-m_N)^{(N-m_N)} }{ (N-m_N)! }}.
 & &
\end{eqnarray*}
 Consequently
%
%
\begin{eqnarray*}
\left ( \prod_{t=1}^{m_N} \frac{e^{-(N-1)\lambda_t}
\left ((N-1)\lambda_t \right )^{z_t}}{z_t!}
\right )^{-1}P \left ( \bm{Z}=\bm{z}\right ) &=& \\
e^{(N-m_N)\left (\sum_{t=1}^{m_N} \lambda_t \right )}
\frac{\left ( (N-m_N)\left ( 1 - \sum_{t=1}^{m_N} \lambda_t \right )\right )^{N-m_N - \sum_{t=1}^{m_N} z_t } (N-m_N)! }{(N-m_N)^{(N-m_N)}
 \left ( N-m_N - \sum_{t=1}^{m_N} z_t \right )!} \times & & \\
\prod_{t=1}^{m_N} \left ( e^{-(m_N-1)\lambda_t}
\left ( \frac{N-m_N}{N-1}\right )^{z_t} \right ) &=& \\
%
%
\underbrace{e^{(N-2m_N+1)\left (\sum_{t=1}^{m_N} \lambda_t \right )}
\left ( 1 - \sum_{t=1}^{m_N} \lambda_t \right )^{N-m_N - \sum_{t=1}^{m_N} z_t}}_{(A)}
\underbrace{\frac{(N-m_N)! }{(N-m_N)^{\sum_{t=1}^{m_N} z_t}
 \left ( N-m_N - \sum_{t=1}^{m_N} z_t \right )!}}_{(B)} \times & & \\
\underbrace{\left ( \frac{N-m_N}{N-1}\right )^{\sum_{t=1}^{m_N} z_t }}_{(C)}. & & 
\end{eqnarray*}
\noindent Regarding (A)
%
%
\begin{eqnarray*}
\left ( \frac{N-m_N}{N-1}\right )^{\sum_{t=1}^{m_N} z_t } &\leq& 1, \\
\left ( \frac{N-m_N}{N-1}\right )^{\sum_{t=1}^{m_N} z_t } &\geq&
\left ( 1 - \frac{m_N}{N} \right )^{b_N} \\
&=&
 \exp \left ( - \frac{b_N m_N}{N} +
 O \left ( b_N \left ( \frac{m_N}{N}\right )^2 \right ) \right ).
\end{eqnarray*}
Hence
%
%
\begin{equation}
\label{eq lemma: bound on the multinomial to poisson pmf ratio first term}
 \left ( \frac{N-m_N}{N-1}\right )^{\sum_{t=1}^{m_N} z_t } =
 \exp \left ( O \left ( \frac{b_N m_N}{N} \right ) \right ).
\end{equation}

\noindent Regarding (B)
%
%
\begin{eqnarray*}
 \frac{(N-m_N)! }{(N-m_N)^{\sum_{t=1}^{m_N} z_t}
 \left ( N-m_N - \sum_{t=1}^{m_N} z_t \right )!} &=&
 \prod_{k=0}^{\sum_{t=1}^{m_N} z_t - 1} \left (1 - \frac{k}{N-m_N} \right ) \\
 &\leq& 1, \\
 \frac{(N-m_N)! }{(N-m_N)^{\sum_{t=1}^{m_N} z_t}
 \left ( N-m_N - \sum_{t=1}^{m_N} z_t \right )!}  &\geq&
\left (1 - \frac{b_N}{N-m_N} \right )^{b_N} \\
&=& \exp \left (  O \left ( \frac{b_N^2}{N}  \right ) \right ).
\end{eqnarray*}
Thus
%
%
\begin{equation}
\label{eq lemma: bound on the multinomial to poisson pmf ratio second term}
 \left ( \frac{N-m_N}{N-1}\right )^{\sum_{t=1}^{m_N} z_t } =
 \exp \left (  O \left ( \frac{b_N^2}{N}  \right ) \right ).
\end{equation}

\noindent Finally, regarding (C)
%
%
\begin{eqnarray*}
\log \left ( e^{ (N-2m_N+1)\left (\sum_{t=1}^{m_N} \lambda_t \right ) }
\left ( 1 - \sum_{t=1}^{m_N} \lambda_t \right )^{ N-m_N - \sum_{t=1}^{m_N} z_t } \right ) &=& \\
 (N-2m_N+1)\left (\sum_{t=1}^{m_N} \lambda_t \right ) +
 \left ( N-m_N - \sum_{t=1}^{m_N} z_t  \right )
 \log \left ( 1 - \sum_{t=1}^{m_N} \lambda_t \right ) &=& \\
 (N-2m_N+1)\left (\sum_{t=1}^{m_N} \lambda_t \right ) +
 \left ( N-m_N - \sum_{t=1}^{m_N} z_t  \right )
 \left ( - \sum_{t=1}^{m_N} \lambda_t  +
 O \left (  \left ( \sum_{t=1}^{m_N} \lambda_t \right )^2 \right ) \right ) &=& \\
 \left ( -m_N+\sum_{t=1}^{m_N} z_t  + 1 \right )
 \left (\sum_{t=1}^{m_N} \lambda_t \right ) + O \left ( \frac{m_N^2}{N} \right ) &=& \\
 O \left ( \frac{m_N \left ( m_N+b_N\right )}{N} \right ) + O \left ( \frac{m_N^2}{N} \right ). & &
\end{eqnarray*}
Hence
\begin{equation}
\label{eq lemma: bound on the multinomial to poisson pmf ratio third term}
e^{ (N-2m_N+1)\left (\sum_{t=1}^{m_N} \lambda_t \right ) }
\left ( 1 - \sum_{t=1}^{m_N} \lambda_t \right )^{ N-m_N - \sum_{t=1}^{m_N} z_t } =
\exp \left ( 
 O \left ( \frac{m_N \max \left ( m_N, b_N\right )}{N} \right )\right ).
\end{equation}

\noindent When $b_N= O(m_N)$,
 Eqs.~(\ref{eq lemma: bound on the multinomial to poisson pmf ratio first term}),
 (\ref{eq lemma: bound on the multinomial to poisson pmf ratio second term}) and
 (\ref{eq lemma: bound on the multinomial to poisson pmf ratio third term})
 imply the desired result.

\hfill Q.E.D.

\vspace{3pt}

%
\begin{lemma} \label{lemma: exponential bound poisson multinomial cdf}
\noindent Let
 $\bm{\lambda}=
  \left [ \lambda_t \right ]_{1 \leq t \leq m_N}=
  \left [ \lambda_N \left ( \bm{v}_t\right ) \right ]_{1 \leq t \leq m_N}$,
%
%
\begin{eqnarray*}
 Z &\sim& \mbox{Poisson} \left ( \left ( N-1\right ) \sum_{t=1}^{m_N} \lambda_t \right ),\\[3pt]
 Z' &\sim& \mbox{Binomial} \left ( N-m_N, \sum_{t=1}^{m_N} \lambda_t \right ).
\end{eqnarray*}
 For any positive $b_N$, we have
%
%
\begin{equation}
 \max \left ( P \left ( Z > e^2 \Lambda m_N \right ),
 P \left ( Z' > e^2 \Lambda m_N \right )  \right ) \leq
 \exp \left ( - e^2 \Lambda m_N \right )
\end{equation}

\end{lemma}


\noindent \underline{Proof}:
Using Markov's inequality, for any positive $b_N$ and $\tau$ we have
%
%
\begin{equation*}
 P \left ( Z > b_N \right ) \leq \phi_Z \left ( \tau \right ) =
 E \left [ \exp \left ( \tau \left ( Z  - b_N \right ) \right ) \right ]
\end{equation*}
 where
$$\phi_Z (\tau) = \exp \left ( \left ( \left ( N-m_N\right ) \sum_{t=1}^{m_N} \lambda_t \right )
  \left ( e^{\tau} - 1 \right ) - \tau b_N\right ),$$
 and $\phi(.)$ is convex.
 Hence
%
%
\begin{equation*}
 P \left ( Z > b_N \right ) \leq \inf_{\tau > 0} \phi \left ( \tau \right ) =
 \phi_Z \left ( \tau^*\right )
\end{equation*}
 where $\tau^*$ is such that $\phi'(\tau^*)=0$; $\phi'(.)$ being the derivative of $\phi()$.
%
%
\begin{eqnarray*}
\tau^* &=&
 \log \left ( \frac{b_N}{ \left ( N-m_N\right )
 \sum_{t=1}^{m_N} \lambda_t }\right ) \\[3pt]
 \phi_Z \left ( \tau^*\right ) &=&
 \exp \left ( -b_N \left ( \left ( \log \left ( \frac{b_N}{\left ( N-m_N\right )
 \sum_{t=1}^{m_N} \lambda_t } \right ) - 1\right ) +
 \left ( N-m_N\right ) \sum_{t=1}^{m_N} \lambda_t \right ) \right ) \\[3pt]
 &\leq&
 \exp \left ( -b_N \left ( \log \left ( \frac{b_N}{\Lambda m_N} \right ) - 1\right )\right )
\end{eqnarray*}
 because $(N-1)\max_{1 \leq t \leq m_N} \lambda_t \leq \Lambda$.
When $b_N= e^2 \Lambda m_N$, we have
%
%
\begin{equation*}
 P \left ( Z > 2\Lambda m_N \right ) \leq \exp \left ( - e^2 \Lambda m_N \right )
\end{equation*}

\noindent In a similar manner
%
%
\begin{equation*}
 P \left ( Z' > b_N \right ) \leq \phi_{Z'} \left ( \tau \right ) =
 E \left [ \exp \left ( \tau \left ( Z'  - b_N \right ) \right ) \right ]
\end{equation*}
 where
%
%
\begin{eqnarray*}
 \phi_{Z'} (\tau) &=&
 \left ( 1 + \left ( \sum_{t=1}^{m_N} \lambda_t \right )
 \left ( e^{\tau} - 1 \right )  \right )^{N-1} e^{-\tau b_N} \\[3pt]
 &\leq&
 \phi_Z (\tau)
\end{eqnarray*}
 where we have used the fact that $1+x \leq e^x$ for any nonnegative $x$.
 Hence
%
%
\begin{eqnarray*}
 P \left ( Z' > b_N \right ) &\leq&  \phi_Z (\tau^*) \\[3pt]
 &\leq&
 \exp \left ( -b_N \left ( \log \left ( \frac{b_N}{\Lambda m_N} \right ) - 1\right )\right )
\end{eqnarray*}
We obtain the desired result when $b_N=e^2 \Lambda m_N$.

\hfill Q.E.D.

\vspace{6pt}

%
\begin{lemma} \label{lemma: second bound characteristic function}

\noindent Let
 $\tilde{\tilde{\bm{n}}}_{M}=\left [ \tilde{\tilde{n}}_{t|M} \right ]_{1 \leq t \leq m_N}$,
 $\tilde{\tilde{\bm{n}}}_{U}=\left [ \tilde{\tilde{n}}_{t|U} \right ]_{1 \leq t \leq m_N}$,
 $\tilde{\tilde{\bm{n}}}=\left [ \tilde{\tilde{n}}_{t}\right ]_{1 \leq t \leq m_N}$,
 where $\tilde{\tilde{n}}_t=\tilde{\tilde{n}}_{t|M}+\tilde{\tilde{n}}_{t|U}$,
 $\tilde{\tilde{\bm{p}}}_N = \left [ \tilde{\tilde{p}}_{Nt} \right ]_{1 \leq t \leq m_N}$, and
 $\tilde{\tilde{\bm{\lambda}}}_N =
 \left [ \tilde{\tilde{\lambda}}_{Nt} \right ]_{1 \leq t \leq m_N}$ be
 such that $\left ( \tilde{\tilde{\bm{p}}}_N,
 \tilde{\tilde{\bm{\lambda}}}_N \right ) \stackrel{d}{=}
 \left ( \bm{p}_N, \bm{\lambda}_N\right )$,  
 $\tilde{\tilde{n}}_{1|M},\ldots,\tilde{\tilde{n}}_{m_N|M}$,
 $\tilde{\tilde{n}}_{1|U},\ldots,\tilde{\tilde{n}}_{m_N|U}$ are
 conditionally independent given $\tilde{\tilde{\bm{p}}}_N$ and
 $\tilde{\tilde{\bm{\lambda}}}_N$ and
%
%
 \begin{eqnarray*}
 \tilde{\tilde{n}}_{t|M} \left |
 \left ( \tilde{\tilde{\bm{p}}}_N, \tilde{\tilde{\bm{\lambda}}}_N \right ) \right .
 &\sim& \mbox{Bernoulli} \left ( \tilde{\tilde{p}}_{Nt} \right ) \\[6pt] 
 \tilde{\tilde{n}}_{t|U} \left |
 \left ( \tilde{\tilde{\bm{p}}}_N, \tilde{\tilde{\bm{\lambda}}}_N\right ) \right .
 &\sim& \mbox{Poisson} \left ( (N-1 ) \tilde{\tilde{\lambda}}_{Nt} \right ).
 \end{eqnarray*}
 Then for any $\bm{T}_N: I\!\!N^{m_N} \rightarrow I\!\!R^d$ and
 $\bm{\omega} \in I\!\!R^d$
%
%
\begin{equation}
 \left |
 E \left [ e^{\jmath \bm{\omega}^{\top} \bm{T}_N \left ( \tilde{\bm{n}} \right ) } \right ] -
 E \left [ e^{\jmath \bm{\omega}^{\top} \bm{T}_N \left ( \tilde{\tilde{\bm{n}}} \right ) }
 \right ] \right | =
O \left ( m_N^2/N\right ),
\end{equation}
 where $\tilde{\bm{n}}$ is as in
 Lemma~\ref{lemma: first bound characteristic function}.

\end{lemma}


\noindent \underline{Proof}: Define the events
%
%
\begin{eqnarray*}
 \tilde{\cal E}_{3N} &=&
 \left \{ \sum_{t=1}^{m_N} \tilde{n}_{t|U} \leq e^2 \Lambda m_N \right \}, \\[3pt]
 {\tilde{\tilde{\cal E}}}_{3N} &=&
 \left \{ \sum_{t=1}^{m_N} \tilde{\tilde{n}}_{t|U} \leq e^2 \Lambda m_N \right \}.
\end{eqnarray*}
 First note that
%
%
\begin{equation}
 E \left [  e^{\jmath \bm{\omega}^{\top} \bm{T}_N \left ( \tilde{\bm{n}} \right ) } \right ] =
 E \left [ I \left ( \tilde{\cal E}_{3N} \right )
 e^{\jmath \bm{\omega}^{\top} \bm{T}_N \left ( \tilde{\bm{n}} \right ) } \right ] +
 P\left ( \tilde{\cal E}_{3N}^c \right ) E \left [ \left .
 e^{\jmath \bm{\omega}^{\top} \bm{T}_N \left ( \tilde{\bm{n}} \right ) } \right |
 \tilde{\cal E}_{3N}^c  \right ]
\end{equation}
Using a change of measure, we have
%
%
\begin{equation}
 E \left [ I \left ( \tilde{\cal E}_{3N} \right )
 e^{\jmath \bm{\omega}^{\top} \bm{T}_N \left ( \tilde{\bm{n}} \right ) } \right ] =
 E \left [ I \left ( {\tilde{\tilde{\cal E}}}_{3N} \right )
 \tilde{\tilde{r}} \left ( \tilde{\tilde{\bm{n}}}_M,
 \tilde{\tilde{\bm{n}}}_U, \tilde{\tilde{\bm{p}}}_N,
 \tilde{\tilde{\bm{\lambda}}}_N \right )
 e^{\jmath \bm{\omega}^{\top} \bm{T}_N \left ( \tilde{\tilde{\bm{n}}} \right ) } \right ],
\end{equation}
 where
%
%
\begin{eqnarray}
 \tilde{\tilde{r}} \left ( \tilde{\tilde{\bm{n}}}_M, \tilde{\tilde{\bm{n}}}_U,
 \tilde{\tilde{\bm{p}}}_N, \tilde{\tilde{\bm{\lambda}}}_N \right ) &=&
 \frac{P \left ( \left . \tilde{\tilde{\bm{N}}}_M=\tilde{\tilde{\bm{n}}}_M,
 \tilde{\tilde{\bm{N}}}_U=\tilde{\tilde{\bm{n}}}_U \right |
 \tilde{\tilde{\bm{P}}}_N=\tilde{\tilde{\bm{p}}}_N,
 \tilde{\tilde{\bm{\Lambda}}}_N=\tilde{\tilde{\bm{\lambda}}}_N \right )}{
 P \left ( \left . \tilde{\bm{N}}_M=\tilde{\tilde{\bm{n}}}_M,
 \tilde{\bm{N}}_U=\tilde{\tilde{\bm{n}}}_U \right |
 \tilde{\bm{P}}_N=\tilde{\tilde{\bm{p}}}_N,
 \tilde{\bm{\Lambda}}_N=\tilde{\tilde{\bm{\lambda}}}_N \right )} \nonumber \\[3pt]
&=&
\frac{P \left ( \left .
 \tilde{\tilde{\bm{N}}}_U=\tilde{\tilde{\bm{n}}}_U \right |
 \tilde{\tilde{\bm{P}}}_N=\tilde{\tilde{\bm{p}}}_N,
 \tilde{\tilde{\bm{\Lambda}}}_N=\tilde{\tilde{\bm{\lambda}}}_N \right )}{
 P \left ( \left .
 \tilde{\bm{N}}_U=\tilde{\tilde{\bm{n}}}_U \right |
 \tilde{\bm{P}}_N=\tilde{\tilde{\bm{p}}}_N,
 \tilde{\bm{\Lambda}}_N=\tilde{\tilde{\bm{\lambda}}}_N \right )}.
\end{eqnarray}
 Hence
%
%
\begin{eqnarray*}
E \left [  e^{\jmath \bm{\omega}^{\top} \bm{T}_N \left ( \tilde{\bm{n}} \right ) } \right ] &=&
 E \left [
 e^{\jmath \bm{\omega}^{\top} \bm{T}_N \left ( \tilde{\tilde{\bm{n}}} \right ) } \right ] -
 E \left [ I \left ( {\tilde{\tilde{\cal E}}}_{3N}^c \right )
 e^{\jmath \bm{\omega}^{\top} \bm{T}_N \left ( \tilde{\tilde{\bm{n}}} \right ) } \right ] +
 \nonumber \\[3pt]
 & &
 E \left [ I \left ( {\tilde{\tilde{\cal E}}}_{3N} \right )
 \left ( \tilde{\tilde{r}} \left ( \tilde{\tilde{\bm{n}}}_M,
 \tilde{\tilde{\bm{n}}}_U, \tilde{\tilde{\bm{p}}}_N,
 \tilde{\tilde{\bm{\lambda}}}_N \right ) -1 \right )
 e^{\jmath \bm{\omega}^{\top} \bm{T}_N \left ( \tilde{\tilde{\bm{n}}} \right ) } \right ] +
\nonumber \\[3pt]
& &
P\left ( \tilde{\cal E}_{3N}^c \right ) E \left [ \left .
 e^{\jmath \bm{\omega}^{\top} \bm{T}_N \left ( \tilde{\bm{n}} \right ) } \right |
\tilde{\cal E}_{3N}^c  \right ].
\end{eqnarray*}
Consequently
%
%
\begin{eqnarray*}
 \left |
 E \left [  e^{\jmath \bm{\omega}^{\top} \bm{T}_N \left ( \tilde{\bm{n}} \right ) } \right ] -
 E \left [ e^{\jmath \bm{\omega}^{\top} \bm{T}_N \left ( \tilde{\tilde{\bm{n}}} \right ) }
 \right ] \right | &\leq&
 P \left ( {\tilde{\tilde{\cal E}}}_{3N}^c \right ) +
 P\left ( {\tilde{\cal E}}_{3N}^c \right ) \nonumber \\[3pt]
 & &
 E \left [ I \left ( {\tilde{\tilde{\cal E}}}_{3N} \right )
 \left | \tilde{\tilde{r}} \left ( \tilde{\tilde{\bm{n}}}_M, \tilde{\tilde{\bm{n}}}_U,
 \tilde{\tilde{\bm{p}}}_N, \tilde{\tilde{\bm{\lambda}}}_N \right ) -1 \right | \right ].
\end{eqnarray*}
From Lemma~(\ref{lemma: bound on the multinomial to poisson pmf ratio}), we have
%
%
\begin{equation*}
I \left ( {\tilde{\tilde{\cal E}}}_{3N} \right )
\left | \tilde{\tilde{r}} \left ( \tilde{\tilde{\bm{n}}}_M, \tilde{\tilde{\bm{n}}}_U,
\tilde{\tilde{\bm{p}}}_N,\tilde{\tilde{\bm{\lambda}}}_N \right ) -1 \right | =
O \left ( m_N^2/N\right ).
\end{equation*}
From Lemma~(\ref{lemma: exponential bound poisson multinomial cdf}), we also have
%
%
\begin{equation*}
 \max \left ( P \left ( {\tilde{\tilde{\cal E}}}_{3N}^c \right ),
 P\left ( {\tilde{\cal E}}_{3N}^c \right ) \right ) \leq
 \exp \left ( - e^2 \Lambda m_N \right ).
\end{equation*}
Therefore
%
%
\begin{eqnarray*}
 \left |
 E \left [  e^{\jmath \bm{\omega}^{\top} \bm{T}_N \left ( \tilde{\bm{n}} \right ) } \right ] -
 E \left [ e^{\jmath \bm{\omega}^{\top} \bm{T}_N \left ( \tilde{\tilde{\bm{n}}} \right ) }
 \right ] \right | =
O \left ( m_N^2/N\right ).
\end{eqnarray*}

\hfill Q.E.D.


\section{Proof of Corollary 1}

\noindent \underline{Proof}:

%
\begin{eqnarray}
\left | E \left [ e^{\jmath \bm{\omega}^{\top} \bm{T}_N \left ( \bm{n} \right ) } \right ] -
 E \left [ e^{\jmath \bm{\omega}^{\top} \bm{\xi}} \right ]
 \right | &\leq&
 \left | E \left [ e^{\jmath \bm{\omega}^{\top} \bm{T}_N \left ( \bm{n} \right ) } \right ] -
 E \left [ e^{\jmath \bm{\omega}^{\top} \bm{T}_N \left ( \tilde{\tilde{\bm{n}}} \right ) } \right ]
 \right | + \nonumber \\[3pt]
 & &
 \left | E \left [ e^{\jmath \bm{\omega}^{\top} \bm{\xi}} \right ] -
 E \left [ e^{\jmath \bm{\omega}^{\top} \bm{T}_N \left ( \tilde{\tilde{\bm{n}}} \right ) } \right ]
 \right | \nonumber \\[3pt]
 &=&
\label{eq corollary asymptotic independence triangle inequality}
 \left | E \left [ e^{\jmath \bm{\omega}^{\top} \bm{\xi}} \right ] -
 E \left [ e^{\jmath \bm{\omega}^{\top} \bm{T}_N \left ( \tilde{\tilde{\bm{n}}} \right ) } \right ]
 \right | + O \left ( m_N^2/N\right )
\end{eqnarray}
 where the last inequation follows from Theorem~\ref{theorem asymptotic independence}.
 Since $\bm{T}_N \left ( \tilde{\tilde{\bm{n}}} \right )
 \stackrel{d}{\rightarrow} \bm{\xi}$, we have
%
%
\begin{equation*}
\lim_{N \rightarrow \infty}
\left | E \left [ e^{\jmath \bm{\omega}^{\top} \bm{\xi}} \right ] -
 E \left [ e^{\jmath \bm{\omega}^{\top} \bm{T}_N \left ( \tilde{\tilde{\bm{n}}} \right ) } \right ]
 \right | = 0
\end{equation*}
Hence Eq.~(\ref{eq corollary asymptotic independence triangle inequality}) implies the following limit
%
%
\begin{equation*}
 \lim_{N \rightarrow \infty}
 \left | E \left [ e^{\jmath \bm{\omega}^{\top} \bm{\xi}} \right ] -
 E \left [ e^{\jmath \bm{\omega}^{\top} \bm{T}_N \left ( \bm{n} \right ) } \right ]
 \right | = 0
\end{equation*}
 and the convergence in distribution of $\bm{T}_N(\bm{n})$ to $\bm{\xi}$ by
 the Continuity theorem \cite[Theorem 26.3]{billingsley_1995}.

\hfill Q.E.D.


\section{E step}

%
\begin{eqnarray}
 P \left ( n_i \big | c_{ig}=1\right ) &=&
 I(n_i=0) (1-p_{(g)}) e^{-\lambda_{(g)}} + \nonumber \\
 & &
 I(n_i>0) \left ( p_{(g)} + (1-p_{(g)}) \frac{\lambda_{(g)}}{n_i}\right ) \frac{e^{-\lambda_{(g)}} \lambda_{(g)}^{n_i-1}}{(n_i-1)!}  \\[6pt]
%
%
\label{eq: E-step 1}
 P \left ( c_{ig}=1 \big | n_i \right ) &=&
 \frac{\alpha_{(g)} P \left ( n_i \big | c_{ig}=1\right )}{\sum_{g'=1}^G \alpha_{(g')} P \left ( n_i \big | c_{ig'}=1\right )} \\[6pt]
%
%
P \left ( n_{i|M} = 1 \big | n_i, c_{ig}=1 \right ) &=&
\frac{p_{(g)} n_i}{p_{(g)} n_i+(1-p_{(g)}) \lambda_{(g)}} \\[6pt]
%
%
 P \left ( n_{i|U} = n_i \big | n_i, c_{ig}=1 \right ) &=&
 I(n_i=0) + I(n_i>0) \frac{(1-p_{(g)}) \lambda_{(g)}}{p_{(g)} n_i+(1-p_{(g)})\lambda_{(g)}} \\[6pt]
%
%
 P \left ( n_{i|U} = n_i-1 \big | n_i, c_{ig}=1 \right ) &=&
 \frac{p_{(g)} n_i}{p_{(g)} n_i+(1-p_{(g)})\lambda_{(g)}} \\[6pt]
%
%
\label{eq: E-step 2}
E \left [ c_{ig} n_{i|M} \big | n_i \right ] &=&
 P \left ( c_{ig}=1 \big | n_i \right ) E \left [ n_{i|M} \big | n_i, c_{ig}=1 \right ] \\[6pt]
%
%
E \left [ n_{i|M} \big | n_i, c_{ig}=1 \right ] &=&
\frac{p_{(g)} n_i}{p_{(g)} n_i+(1-p_{(g)}) \lambda_{(g)}} \\[6pt]
%
%
%
\label{eq: E-step 3}
E \left [ c_{ig} n_{i|U} \big | n_i \right ] &=&
 P \left ( c_{ig}=1 \big | n_i \right ) E \left [ n_{i|U} \big | n_i, c_{ig}=1 \right ] \\[6pt]
%
%
E \left [ n_{i|U} \big | n_i, c_{ig}=1 \right ] &=&
\left ( \frac{p_{(g)}(n_i-1)+(1-p_{(g)})\lambda_{(g)}}{p_{(g)} n_i+(1-p_{(g)})\lambda_{(g)}} \right ) n_i
\end{eqnarray}


\section{SAS code}

{\footnotesize
\begin{lstlisting}[breaklines]

libname local 'some_folder\data';

options symbolgen;
proc printto log='some_folder\log\empirical_log.txt' new;
run;
proc printto print='some_folder\output\empirical_out.txt' new;
run;

/*----------------------------------------------*/
/*----------------------------------------------*/
/* Macro variables
/*----------------------------------------------*/

%let first_estimates = 1;
%let min_num_classes = 16;
%let max_num_classes = 16;		/* ideally between 5 and 10 */

%let init_p = 0.9;

%let max_num_iter	 = 100000;	/* ideally no less than 2000 */

/* Bootstrap */
%let num_boot_samples	= 0;	/* ideally 1000 */
%let boot_num_iter	 	= 1;

/* Parametric bootstrap */
%let pb_num_samples	= 0;	/* ideally 100 or 1000 */
%let pb_num_iter	= 1;
%let conf_level		= 0.5;	/* ideally 0.05 */

/* Seed to select the subsample */
%let streaminit_seed = 200;

/* Initial values */
%let num_init_values = 1;		/* ideally no less than 10 */

/* */
%let log_eps = 1e-30;

%let init_t=0.5;
%let init_step=0.1;

/* */
%let rel_diff=0.1;
%let history_length=100;
%let EPS=1e-6;

/*----------------------------------------------*/
/*----------------------------------------------*/
/* Macro to select the sample
/* n_1,...,n_{m_N}
/*----------------------------------------------*/

%macro select_sample(in_dset=,out_dset=,target_size=,streaminit_seed=,method=);

proc sql;
 create table ss_target_ssize as
 select &target_size. as target_size, sum(freq) as pop_size
 from &in_dset.;
quit;

/* */
proc sql;
 create table ss_subsample_0 as
 select a.*, b.target_size, b.pop_size
 from &in_dset. a, ss_target_ssize b;
quit;

/* */
%if &method.=0 %then %do;

data &out_dset.;
 set ss_subsample_0(rename=(freq=old_freq));
 retain current_sample_size current_pop_size current_target_size;
 call streaminit(&streaminit_seed.);
 if _n_ eq 1 then do;
  freq=RAND('HYPER',pop_size,old_freq,target_size);
  current_sample_size=freq;
  current_pop_size=pop_size-old_freq;
  current_target_size=target_size-freq;
 end;
 else do;
  if current_target_size gt 0 then do;
   freq=RAND('HYPER',current_pop_size,old_freq,current_target_size);
   current_sample_size=current_sample_size+freq;
   current_pop_size=current_pop_size-old_freq;
   current_target_size=current_target_size-freq;
  end;
  else do;
   freq=0;
  end;
 end;

 if freq gt 0;
 keep n_i freq;
run;

 %end;
 %else %do;

 data &out_dset.;
  set ss_subsample_0(rename=(freq=old_freq));
  retain current_sample_size sum_old_freq;
  current_sample_size=0;
  sum_old_freq=0;
  if current_sample_size lt pop_size then do;
   freq=rand('BINOMIAL',old_freq/(pop_size-sum_old_freq),pop_size-current_sample_size);
   current_sample_size=current_sample_size+freq;
   sum_old_freq=sum_old_freq+old_freq;
  end;

 if freq gt 0;
 keep n_i freq;
run;

%end;

%mend select_sample;

/*-------------------------------------*/
/*-------------------------------------*/
/* select_initial_t()
/*-------------------------------------*/

%macro select_initial_t(in_sample=,in_t=,out_initial_values=);

%local num_classes g;

proc sql;
 create table max_n_i as
 select max(n_i) as max_n_i, sum(freq) as total_freq
 from &in_sample.;
quit;

data first_lambda;
 set max_n_i;
 lambda_g=(&in_t.)**2;
 n_i_lb=max(0,lambda_g-&in_t.*sqrt(lambda_g));
 n_i_ub=lambda_g+&in_t.*sqrt(lambda_g);
run;

data remaining_lambdas;
 set first_lambda;
 retain lambda_g n_i_lb n_i_ub;
 do while(n_i_ub lt max_n_i);
  lambda_g=(&in_t.+sqrt((&in_t.)**2+4*n_i_ub))**2/4;
  n_i_lb=lambda_g-&in_t.*sqrt(lambda_g);
  n_i_ub=lambda_g+&in_t.*sqrt(lambda_g);
  output;
 end;
run;

data all_lambdas;
 set first_lambda remaining_lambdas;
 drop max_n_i;
run;

proc sql;
 create table lambdas_and_freqs as
 select a.*,b.*
 from &in_sample. a, all_lambdas b;
quit;

proc sql;
 create table lambda_freq as
 select lambda_g, sum((n_i ge n_i_lb)*(n_i lt n_i_ub)*freq) as freq, mean(total_freq) as total_freq
 from lambdas_and_freqs group by lambda_g;
quit;

data initial_values_in_rows;
 set lambda_freq(where=(freq gt 0));
 g=_n_;
 alpha_g=freq/total_freq;
 p_g=&init_p.;
 keep g lambda_g alpha_g p_g;
run;

proc sql;
 select count(*) into:num_classes from initial_values_in_rows;
quit;
proc sql;
 create table &out_initial_values. as
 select &in_t. as t, &num_classes. as num_classes, %do g=1 %to &num_classes.; %if &g.>1  %then %do; , %end; sum((g=&g.)*alpha_g) as alpha_&g., sum((g=&g.)*p_g) as p_&g., sum((g=&g.)*lambda_g) as lambda_&g.  %end;
 from initial_values_in_rows;
quit;

%mend select_initial_t;

/*-------------------------------------*/
/*-------------------------------------*/
/* select_initial()
/*-------------------------------------*/

%macro select_initial(in_sample=,in_initial_t=,in_initial_step=,in_num_classes=,out_initial_values=);

%local num_classes t;

%select_initial_t(in_sample=&in_sample.,in_t=&in_initial_t.,out_initial_values=out_initial_values);

proc sql;
 select num_classes into:num_classes from out_initial_values;
quit;

data params;
 t=&in_initial_t.;
 step=&in_initial_step.;
 output;
run;

%if &in_num_classes.=&num_classes. %then %do;
 data &out_initial_values.;
 	set out_initial_values;
	drop t num_classes;
 run;
%end;
%else %if &in_num_classes.<&num_classes. %then %do;
 /* must increase t */
 %do %while(&in_num_classes.~=&num_classes.);
	%if &in_num_classes.>&num_classes.%then %do;
	 data new_params;
	 	set params(rename=(t=old_t step=old_step));
		t=old_t*exp(-old_step/2);
		step=old_step/2;
		keep t step;
	 run;
	%end;
	%else %do;
	 data new_params;
	 	set params(rename=(t=old_t));
		t=old_t*exp(step);
		keep t step;
	 run;
	%end;
	data params;
	  set new_params;
	run;
	proc sql;
  		select t into:t from params;
 	quit;
 	%select_initial_t(in_sample=&in_sample.,in_t=&t.,out_initial_values=out_initial_values);
 	proc sql;
  		select num_classes into:num_classes from out_initial_values;
 	quit;
 %end;
 data &out_initial_values.;
 	set out_initial_values;
	drop t num_classes;
 run;
%end;
%else %do;
 /* must decrease t */
  %do %while(&in_num_classes.~=&num_classes.);
	%if &in_num_classes.<&num_classes.%then %do;
	 data new_params;
	 	set params(rename=(t=old_t step=old_step));
		t=old_t*exp(old_step/2);
		step=old_step/2;
		keep t step;
	 run;
	%end;
	%else %do;
	 data new_params;
	 	set params(rename=(t=old_t));
		t=old_t*exp(-step);
		keep t step;
	 run;
	%end;
	data params;
	  set new_params;
	run;
	proc sql;
  		select t into:t from params;
 	quit;
 	%select_initial_t(in_sample=&in_sample.,in_t=&t.,out_initial_values=out_initial_values);
 	proc sql;
  		select num_classes into:num_classes from out_initial_values;
 	quit;
 %end;
 data &out_initial_values.;
 	set out_initial_values;
	drop t num_classes;
 run;
%end;

%mend select_initial;

/*----------------------------------------------*/
/*----------------------------------------------*/
/* Stopping criterion
/*----------------------------------------------*/

%macro stopping_criterion(num_classes=,iteration_history=,in_current_iter=,out_dset=);

%local k num_missing;

proc sql;
 select count(*) into:history_obs from &iteration_history.;
quit;

data all_estimates;
 set &iteration_history.(rename=(parameter=param value=estimate));
run;

proc sql;
 create table min_max_est as
 select param, k, min(estimate) as min_est, max(estimate) as max_est
 from all_estimates group by param, k;
quit;

data local.min_max_est;
 set min_max_est;
run;

proc sql;
 create table &out_dset. as
 select &in_current_iter. as iter, min((abs(min_est-max_est)<&rel_diff.*max(abs(min_est),&EPS.))) as result,
 max(abs(min_est-max_est)/(abs(min_est)+&EPS.)) as max_rel_diff
 from min_max_est;
quit;


%mend stopping_criterion;

/*----------------------------------------------*/
/*----------------------------------------------*/
/* em_asymptotic()
/*----------------------------------------------*/

%macro em_asymptotic(	in_sample=,
				in_init_values=,
				in_num_iter=,
				out_estimates=,
				num_classes=,
				start_index=,
				boot_index=,
				pb_iter=);

%local k iter conv_result firstobs lastobs next_check_iter;

data estimates_M_step;
 set &in_init_values.;
run;

%let iter=1;
%let conv_result=0;
%let next_check_iter=%eval(&history_length+1);

%do %until(&iter.>&max_num_iter. | &conv_result.=1);

 proc sql;
  create table sample_E_step_0 as
  select a.*,b.*
  from &in_sample.(keep=n_i freq) as a, estimates_M_step as b;
 quit;

data sample_E_step_1;
 set sample_E_step_0;
 if n_i eq 0 then do;
  %do k=1 %to &num_classes.;
   p_n_i_&k.=(1-p_&k.)*exp(-lambda_&k.);
  %end;
 end;
 else do;
  %do k=1 %to &num_classes.;
   p_n_i_&k.=(p_&k.+(1-p_&k.)*lambda_&k./n_i)*PDF('POISSON',n_i-1,lambda_&k.);
  %end;
 end;

 %if &num_classes.=1 %then %do;
  total=alpha_1*p_n_i_1;
 %end;
 %else %do;
  total=alpha_1*p_n_i_1 + %do k=2 %to &num_classes.; + alpha_&k.*p_n_i_&k. %end;;
 %end;

  %do k=1 %to &num_classes.;
   E_c_&k.=alpha_&k.*p_n_i_&k./total;
   if p_&k. eq 1 then do;
   	E_n_i_1_&k.=1;
   	E_n_i_2_&k.=(n_i>0)*(n_i-1);
   end;
   else do;
   	E_n_i_1_&k.=n_i*p_&k./(n_i*p_&k.+(1-p_&k.)*lambda_&k.);
   	E_n_i_2_&k.=n_i*((n_i-1)*p_&k.+(1-p_&k.)*lambda_&k.)/(n_i*p_&k.+(1-p_&k.)*lambda_&k.);
   end;
  %end;
 run;

 %if &num_classes.=1 %then %do;
  proc sql;
   create table estimates_M_step as
   select sum(freq*E_c_1)/sum(freq) as alpha_1, sum(freq*E_c_1*E_n_i_1_1)/sum(freq*E_c_1) as p_1,
   sum(freq*E_c_1*E_n_i_2_1)/sum(freq*E_c_1) as lambda_1
   from sample_E_step_1;
  quit;
 %end;
 %else %do;
  proc sql;
   create table estimates_M_step as
    select %do k=1 %to &num_classes.;sum(freq*E_c_&k.)/sum(freq) as alpha_&k.,
	sum(freq*E_c_&k.*E_n_i_1_&k.)/sum(freq*E_c_&k.) as p_&k.,
	sum(freq*E_c_&k.*E_n_i_2_&k.)/sum(freq*E_c_&k.) as lambda_&k.
 	%if &k.<&num_classes. %then %do; , %end; %end;
   from sample_E_step_1;
  quit;
 %end;

  /* Compute the log-likelihood */
 proc sql;
  create table obs_and_estimates as
  select a.n_i, a.freq, b.*
  from &in_sample.(keep=n_i freq) a,
  estimates_M_step(keep=%do k=1 %to &num_classes.; alpha_&k. p_&k. lambda_&k. %end;) b;
 quit;

 proc sql;
  select count(*) into:num_obs from obs_and_estimates;
 quit;

 /* */
 data mles_and_ll;
  set obs_and_estimates;
  retain total_ll;
  iter=&iter.;
  start_index=&start_index.;
  if(_n_ eq 1) then total_ll=0;
  ll_value=log(&log_eps.+%do k=1 %to &num_classes.;
  %if &k.>1 %then %do; + %end;
  alpha_&k.*((n_i=0)*(1-p_&k.)*exp(-lambda_&k.)+
  (n_i>0)*(p_&k.*PDF('Poisson',n_i-1,lambda_&k.)+(1-p_&k.)*PDF('Poisson',n_i,lambda_&k.))) %end;);
  total_ll=total_ll+freq*ll_value;
  if _n_ eq &num_obs.;
  drop n_i freq ll_value;
 run;

  /* Add to the iteration history */
 data alpha_estimates;
  set mles_and_ll;
  parameter='alpha';
  num_classes=&num_classes.;
  iter=&iter.;
  boot_index=&boot_index.;
  start_index=&start_index.;
  pb_iter=&pb_iter.;
  %do k=1 %to &num_classes.;
   value=alpha_&k.;
   k=&k.;
   output;
  %end;
  keep value parameter num_classes k iter boot_index start_index pb_iter;
 run;

 data p_estimates;
  set mles_and_ll;
  parameter='p';
  num_classes=&num_classes.;
  iter=&iter.;
  boot_index=&boot_index.;
  start_index=&start_index.;
  pb_iter=&pb_iter.;
  %do k=1 %to &num_classes.;
   value=p_&k.;
   k=&k.;
   output;
  %end;
  keep value parameter num_classes k iter boot_index start_index pb_iter;
 run;

 data lambda_estimates;
  set mles_and_ll;
  parameter='lambda';
  num_classes=&num_classes.;
  iter=&iter.;
  boot_index=&boot_index.;
  start_index=&start_index.;
  pb_iter=&pb_iter.;
  %do k=1 %to &num_classes.;
   value=lambda_&k.;
   k=&k.;
   output;
  %end;
  keep value parameter num_classes k iter boot_index start_index pb_iter;
 run;

 data total_ll_estimate;
  set mles_and_ll(rename=(total_ll=value));
  parameter='total_ll';
  num_classes=&num_classes.;
  iter=&iter.;
  boot_index=&boot_index.;
  start_index=&start_index.;
  pb_iter=&pb_iter.;
  keep value parameter num_classes iter boot_index start_index pb_iter;
 run;

 data current_estimates;
  set total_ll_estimate alpha_estimates p_estimates lambda_estimates;
 run;

 %if &iter.=1 %then %do;
   data em_history;
    set current_estimates;
   run;
 %end;
 %else %do;
   proc append base=em_history data=current_estimates;
   run;
 %end;

 data local.em_history;
  set em_history;
 run;

 %if &iter.=&next_check_iter. %then %do;
  %let firstobs=%eval((&iter-&history_length)*(3*&num_classes+1));
  %let lastobs=%eval(&iter*(3*&num_classes+1));
  data in_history;
   set em_history(firstobs=&firstobs. obs=&lastobs. keep=iter parameter value k);
   if parameter ne 'total_ll' and (parameter ne 'alpha' or k ne &num_classes.); 
  run;
  %stopping_criterion(num_classes=&num_classes.,iteration_history=in_history,in_current_iter=&iter.,out_dset=conv_diag);
  proc sql;
   select result into:conv_result from conv_diag;
  quit;
  data local.conv_diag;
   set conv_diag;
  run;
  %let next_check_iter=%eval(&iter+&history_length);
 %end;

 %let iter=%eval(&iter+1);
   /* %do iter=1 %to &num_iter.; */
 %end;

 %if &first_estimates.=1 %then %do;
   data local.iteration_history;
    set em_history;
   run;
   %let first_estimates=0;
 %end;
 %else %do;
   proc append base=local.iteration_history data=em_history;
   run;
 %end;

 data &out_estimates.;
   set mles_and_ll;
   conv_result=&conv_result.;
 run;

 /*----------------------------------*/
 /*----------------------------------*/
%mend em_asymptotic;

/*----------------------------------*/
/*----------------------------------*/
/* Compute LRT critical levels
/* through a parametric boostrap
/*----------------------------------*/

%macro estimate_pb_level(num_classes=,in_mle=,sample_size=,alpha_level=,out_level=,boot_index=);

%local k t;

 data mixture_params;
  set &in_mle.;
  keep %do k=1 %to &num_classes.; alpha_&k. p_&k. lambda_&k. %end;;
 run;

 %do t=1 %to &pb_num_samples.;

  data iid_sample;
   set mixture_params;
   call streaminit(&t.+&num_classes.);
   do i=1 to &sample_size.;
     u=RAND('UNIFORM');
	 class=1 %do k=2 %to &num_classes.; + (u gt (alpha_1 %do l=2 %to &k.; + alpha_&l. %end;)) %end;;
     n_i1=(class=1)*RAND('BERNOULLI',p_1) %do k=2 %to &num_classes.; + (class=&k.)*RAND('BERNOULLI',p_&k.) %end;;
	 n_i2=(class=1)*RAND('POISSON',lambda_1) %do k=2 %to &num_classes.; + (class=&k.)*RAND('POISSON',lambda_&k.) %end;;
	 n_i=n_i1+n_i2;
	 output;
   end;
   keep n_i;
  run;

  proc sql;
   create table pb_sample as
   select n_i, count(*) as freq
   from iid_sample group by n_i;
  quit;

%select_initial(in_sample=pb_sample,
			in_initial_t=&init_t.,
			in_initial_step=&init_step.,
			in_num_classes=&num_classes.,
			out_initial_values=init_values);

 %em_asymptotic(in_sample=pb_sample,
			in_init_values=init_values,
			in_num_iter=&pb_num_iter.,
			out_estimates=pb_mle&num_classes.,
			num_classes=&num_classes.,
			start_index=1,
			boot_index=&boot_index.,
			pb_iter=&t.);

%select_initial(in_sample=pb_sample,
			in_initial_t=&init_t.,
			in_initial_step=&init_step.,
			in_num_classes=&max_num_classes.,
			out_initial_values=init_values);

%em_asymptotic(	in_sample=pb_sample,
 			in_init_values=init_values,
			in_num_iter=&pb_num_iter.,
			out_estimates=pb_mle&max_num_classes.,
			num_classes=&max_num_classes.,
			start_index=1,
			boot_index=&boot_index.,
			pb_iter=&t.);

	proc sql;
	 create table new_stat as
	 select -2*(a.total_ll - b.total_ll) as test_stat
	 from pb_mle&num_classes. a, pb_mle&max_num_classes. b;
	quit;

	%if &t.=1 %then %do;
	 data all_stats&num_classes.;
	  set new_stat;
	 run;
	%end;
	%else %do;
	 proc append base=all_stats&num_classes. data=new_stat;
	 run;
	%end;
%end;

proc sort data=all_stats&num_classes.;
 by test_stat;
run;

data &out_level.;
 set all_stats&num_classes.;
 if _n_ eq max(1,floor(&pb_num_samples.*(1-&alpha_level.)));
run;

%mend estimate_pb_level;

/*----------------------------------*/
/*----------------------------------*/
/* Estimate the parameters
/* Find all the point estimates and
/* the CIs
/*----------------------------------*/

%macro find_all_estimates();

 proc sql;
  select sum(freq) into:target_size  from original_sample;
 quit;

 %local boot_index start_index g current_seed;

%let sample_size=&target_size;

%do g=&min_num_classes. %to &max_num_classes.;

 %do boot_index=0 %to &num_boot_samples.;

  %if &boot_index=0 %then %do;

   %do start_index=1 %to &num_init_values.;

    %let current_seed=%eval(&streaminit_seed+&boot_index+&start_index+&g.);

	%select_initial(in_sample=original_sample,
			in_initial_t=&init_t.,
			in_initial_step=&init_step.,
				in_num_classes=&g.,
				out_initial_values=init_values);

	%em_asymptotic(	in_sample=original_sample,
 				in_init_values=init_values,
				in_num_iter=&max_num_iter.,
				out_estimates=out_estimates,
				num_classes=&g.,
				boot_index=&boot_index.,
				start_index=&start_index.,
				pb_iter=-1);

    %if &start_index.=1 %then %do;
     data estimates_all_starts;
      set out_estimates;
     run;
    %end;
    %else %do;
     proc append base=estimates_all_starts data=out_estimates;
     run;
    %end;

   %end; /* %do start_index=1 */

  /*----------------------------------*/
  /* Select the estimate with the
  /* maximum likelihood
  /*----------------------------------*/
  proc sort data=estimates_all_starts;
  	by descending total_ll;
  run;
  data mle_&g.;
   set estimates_all_starts(obs=1);
 run;
 data new_mles;
	   set mle_&g.;
	   num_classes=&g.;
	   boot_index=&boot_index.;
	   %if &g.=1 %then %do;
   		 overall_p=p_1;
   		overall_lambda=lambda_1;
  	  %end;
  	  %else %do;
   	   overall_p=alpha_1*p_1 %do k=2 %to &g.; + alpha_&k.*p_&k. %end;;
   	   overall_lambda=alpha_1*lambda_1 %do k=2 %to &g.; + alpha_&k.*lambda_&k. %end;;
  	  %end;
 run;
 %if &g.=&min_num_classes. %then %do;
  	  data local.all_mles;
	   set new_mles(keep=num_classes boot_index overall_p overall_lambda conv_result);
	  run;
 %end;
 %else %do;
      proc append base=local.all_mles data=new_mles(keep=num_classes boot_index overall_p overall_lambda conv_result);
      run; 
 %end;

 %if &pb_num_samples.>=1 %then %do;
  %if &g.<&max_num_classes. %then %do;
  /*
  %put *---------------------------------------------------*;
  %put *---------------------------------------------------*;
  %put *---------------------------------------------------*;
  %put * COMPUTING PB_LEVEL FOR G=&g;
  %put *---------------------------------------------------*;
  */

  /* Estimate the correct critical level via a
	 parametric bootstrap procedure */
	%estimate_pb_level(	num_classes=&g.,
					in_mle=mle_&g.,
					sample_size=&sample_size.,
					alpha_level=&conf_level.,
					out_level=pb_level_&g.,
					boot_index=&boot_index.);

   /* Data set to store the mle and level */
   proc sql;
    create table mle_and_level_&g. as
    select a.*, b.test_stat as pb_level, &g. as g
    from mle_&g. a, pb_level_&g. b;
   quit;
 %end;
 %else %do;
   data mle_and_level_&g.;
    set mle_&g.;
	pb_level=.;
    g=&g.;
   run;
 %end;
 %end;

 %end; /* %if &boot_index=0 */
 %else %do;
   /*
  %put *---------------------------------------------------*;
  %put *---------------------------------------------------*;
  %put *---------------------------------------------------*;
  %put * Bootstrap samples
  %put *---------------------------------------------------*;
  */
	%let current_seed=%eval(&streaminit_seed+&boot_index+&start_index+&g.); 

	%select_sample(	in_dset=original_sample,
				out_dset=b_sample&g.,
				target_size=&target_size.,
				streaminit_seed=&current_seed.,
				method=1);

	%select_initial(in_sample=b_sample&g.,
			in_initial_t=&init_t.,
			in_initial_step=&init_step.,
				in_num_classes=&g.,
				out_initial_values=init_values);

	%em_asymptotic(	in_sample=b_sample&g.,
 				in_init_values=init_values,
				in_num_iter=&boot_num_iter.,
				out_estimates=b_mle_&g.,
				num_classes=&g.,
				boot_index=&boot_index.,
				start_index=1,
				pb_iter=-1);

	data new_mles;
	   set b_mle_&g.;
	   num_classes=&g.;
	   boot_index=&boot_index.;
	   %if &g.=1 %then %do;
   		overall_p=p_1;
   		overall_lambda=lambda_1;
  	   %end;
  	   %else %do;
        overall_p=alpha_1*p_1 %do k=2 %to &g.; + alpha_&k.*p_&k. %end;;
        overall_lambda=alpha_1*lambda_1 %do k=2 %to &g.; + alpha_&k.*lambda_&k. %end;;
       %end;
	run;

	proc append base=local.all_mles data=new_mles(keep=num_classes boot_index overall_p overall_lambda conv_result);
	run;
 
 %end;
 %end; /* boot_index */

/* %end; */ /*  %if &g.=&target_class. | &g.=&max_num_classes. %then %do; */

%end; /* %do g=1 %to &max_num_classes.; */

/*Compute the bootstrap confidence intervals*/
proc sort data=local.all_mles;
 by num_classes;
run;

%if &num_boot_samples.>=1 %then %do;

proc sql;
 create table local.all_boot_variances as
 select num_classes, var(overall_p) as var_p, var(overall_lambda) as var_lambda
 from local.all_mles(where=(boot_index gt 0)) group by num_classes;
quit;

proc univariate data=local.all_mles(keep=overall_lambda num_classes) noprint;
 by num_classes notsorted;
 var overall_lambda;
 output out=lambda_b_ci(keep=num_classes P97_5 P2_5) pctlpts=97.5 2.5  pctlpre=P;
run;

proc univariate data=local.all_mles(keep=overall_p num_classes) noprint;
 by num_classes notsorted;
 var overall_p;
 output out=p_b_ci pctlpts=97.5 2.5  pctlpre=P;
run;

data local.all_bootstrap_cis;
 set p_b_ci lambda_b_ci;
 param='lambda';
 if _n_ le 2 then param='p';
run;
%end;

%if &pb_num_samples.>=1 %then %do;

data mles_and_levels;
 set %do g=&min_num_classes. %to &max_num_classes.; mle_and_level_&g. %end;;
 df=3*(&max_num_classes.-g);
 if df gt 0 then standard_level=QUANTILE('CHISQ',1-&conf_level.,df);
 else standard_level=.;
run;

proc sort data=mles_and_levels;
 by descending g;
run;

data local.all_tests;
 set mles_and_levels;
 retain total_ll_max_g;
 if _n_ eq 1 then total_ll_max_g=total_ll;
  test_stat=-2*(total_ll-total_ll_max_g);
  reject_h0_standard=(test_stat>standard_level);
  reject_h0_pb=(test_stat>pb_level);
 if g < &max_num_classes.;
 keep g test_stat df standard_level pb_level reject_h0_standard reject_h0_pb;
run;

%end;

%mend find_all_estimates;

/*----------------------------------*/
/*----------------------------------*/
/* Estimate the parameters
/*----------------------------------*/

%find_all_estimates();

proc printto;
run;

\end{lstlisting}}

\bibliographystyle{Chicago}

\bibliography{asymptotic_rl_errors}
\end{document}